# Numerical studies on steady interaction of low enthalpy hypersonic double wedge flows using different gas models

Qin Li[a*†], Y Wang[b†], Yihui Weng[a], Yunchuan Wu[a], Mengyu Wang[a], Pan Yan[a], Linsen Zhang[b], Wei Su[b]

a. School of Aerospace Engineering, Xiamen University, China, 361102
b. China Academy of Launch Vehicle Technology, Beijing, China, 100076

**Abstract:** Numerical investigations and analyses are carried out on the interactions of low enthalpy hypersonic 30–55° double wedge configuration, particularly focusing on steady cases at conditions similar to the experimental setup by Swantek & Austin [AIAA 2012-284], with $Ma = 7$ and $h_0 = 2.1 MJ/kg$. To achieve a steady solution, Reynolds numbers ($Re$) lower than those in the experiment are used. For increased accuracy, a third-order scheme WENO3-PRM$_{1,1}^2$ [Li et al., J. Sci. Comput., 88(3) (2021) 75-130] with improved resolution is employed. Meanwhile, three gas models, i.e., the perfect, equilibrium, and non-equilibrium gas models, are used to analyze the difference potentials that arise from the physical model. After validating the methods, grid convergence studies are first conducted at $Ma = 7$ and $Re = 2.5 \times 10^5/m$, to determine the appropriate grid resolution for the main computations. Subsequently, comprehensive numerical studies are carried out on the steady interactions and their evolution at $Ma = 7$ and $h_0 = 2.1 MJ/kg$. Specifically: (a) The upper limits of $Re$ are identified where the flows remain steady, and the corresponding interaction characteristics as well as differences in the three gas models are investigated qualitatively and quantitatively, e.g., the shock system, vortex structures, distributions such as pressure, Mach number, and specific heat ratio. Notably, a quasi-normal shock wave is observed within the slip line passage in the case of the perfect gas model. (b) The flow characteristics of the three models, including the interaction pattern, geometric features of triple points, impingements, and separation zone, are studied and compared for $Re = 4, 3,$ and $2 \times 10^4/m$. Differences primarily emerge between the results of the perfect gas model and the real gas model. Specifically, a transmitted shock reflecting above the separation zone is observed in the case of the perfect gas model. The effect of the gas model on temperature and specific heat ratio distributions, as well as the heat transfer and pressure coefficients over the wedge surface are investigated. For an in-depth understanding, the shock polar method is applied for comparison with computational results, while a 1D flow model is proposed to explain the occurrence of the quasi-normal shock wave. Consequently, overall reasonable agreements are achieved. Finally, the effects of variations in Mach number and enthalpy are determined, by alternatively varying the two parameters around $Ma = 7$ and $h_0 = 2.1 MJ/kg$ at $Re = 4 \times 10^4/m$, focusing on alterations in interaction characteristics, thermodynamic properties, and aerodynamic performance.

## 1 Introduction

As a commonly used component, the double wedge is applied in supersonic and hypersonic inlets and fuselages, e.g., as a control surface to generate the moment needed to change the attitude of high-speed vehicles. However, due to the presence of shock waves and intensive interactions, accurately predicting the aerodynamic load, which is critical for aircraft design, such as the peak heat transfer in hypersonic cases, is challenging and essential to understanding the flow mechanism.

---

[*] Corresponding author, email: qin-li@vip.tom.com
[†] Qin Li and Yonghai Wang are co-first authors

To assess the predictive ability of computational fluid dynamics (CFD) for lower hypersonic flows with low to modest stagnation enthalpies [1], a series of investigations was conducted by NATO STO AVT task group 205. Among them, a double wedge configuration was studied which was proposed by Swantek and Austin [2, 3], with experimental results serving as the reference. The double wedge had fore and aft angles of $\theta_1 = 30°$ and $\theta_2 = 55°$, respectively, with corresponding surface lengths of $L_1 = 50.8$ mm and $L_2 = 25.4$ mm. In addition, the width was $L_Z = 101.6$ mm, and a horizontal extension was connected to the aft wedge with $L_3 = 10.82$ mm. The analysis [2, 3] revealed that $L_z/\delta \approx 125$, exceeding the criteria of 85 [4] ($\delta$ is the boundary layer thickness before separation on the fore wedge). Therefore, the flow was perceived to be two-dimensional experimentally. The experimental setup included a range of Mach numbers ($Ma$) and Reynolds numbers ($Re$) at both low and relatively high stagnation enthalpies $h_0$. In hypersonic cases with $Ma = 7$, the unit $Re$ values were relatively low at $1.1 \times 10^6$/m and $4.35 \times 10^5$/m, indicating a laminar flow profile.

In the hypersonic case with a higher $h_0$ of 8 MJ/kg or M7_8 [2, 3], thermochemical non-equilibrium phenomena, e.g., the excitation of molecular vibrations, energy exchange between translation-rotation and vibration, and chemical reactions with finite rates, were considered and their corresponding effects were numerically investigated. These practices were initially summarized in [1] and subsequent advancements were reported in [5-9]. Conversely, the hypersonic case with a lower $h_0$ of 2.1 MJ/kg, has also received interest from the scientific community and is the focus of this study. By convention, this case is abbreviated as M7_2.

Although M7_2 appears to have lower thermal and chemical complexity, achieving consistency between computational and experimental results is challenging. In [1], the results of various numerical investigations, e.g., those of Knight et al., Candler et al., and Celik et al., were analyzed and compared with experimental results. Two different moments in the computations were chosen for comparison purposes: a short one lasting 327 or 270 $\mu s$ and a longer one on the order of $ms$. The short moment referenced the same experimental test time or the duration of schlieren images. The comparisons at the "short time" [1] revealed that while the predictions usually showed larger separation with unexpected heat flux oscillations and different shock structures, the peak value on the aft wedge was close to that of the experiment. However, the flow field was still evolving and far from convergence. At the "long time" [1], the computations predicted oversized separation and lower heat flux on the aft wedge at the interaction locations compared to the experimental data. Subsequent extensive studies on M7_2 utilized thermochemical non-equilibrium gas models [6, 9, 10] rather than perfect gas models [11-15]. Additionally, studies incorporating 3D configurations [6, 13] were also performed. In 2D investigations, some authors calculated the time-averaged heat flux over a chosen period, e.g., (150-320)$\mu s$ in [12, 13], (50-327) $\mu s$ in [10], and (50-320)$\mu s$ in [15], to compare with experimental results at the reference moment of $270\mu s$. While some agreements were reached in these investigations, Ninni et al. and Durna and Celik [9, 12] noted that follow-up flows continue to evolve, with the transmitted shock moving back and forth around the expansion corner [9]. Conversely, Kumar and De [15] argued that the reported unsteadiness in [9] stemmed from insufficient grids and that the flow would become quasi-steady with denser grids. However, the converged interaction pattern in [15] still differed significantly from experimental results [2, 3]. The question arises whether the inconsistency may be attributed to 3D effects. Reinert et al. [6] addressed this by selecting a 3D double wedge and comparing predictions along the center line at $215\mu s$ as well as the average over (3.86-9.03) *ms*. The comparison showed that the location and value of the instant peak heat flux did not match well with the experimental values, with the averaged value indicating earlier separation and a smaller peak heat with a delayed location. Durna and Celik [13] also investigated the averaged center line heat flux using 3D computations over (150-310)$\mu s$, which showed a similar peak distribution but with lesser separation. In terms of gas models, Ninni et al. and Expósito and Rana [9, 10] compared the results using non-equilibrium models to those of perfect gas models in other studies and found no clear distinctions.

In addition to prediction precision, researchers have also focused on studying the flow mechanisms of M7_2. One key area of interest is the interaction pattern. Edney type IV and V interactions are considered as canonical examples of shock–shock interactions, where locally subsonic and supersonic jets are observed during the development and establishment of the flow field [6, 7, 11-15]. However, due to differences in solving the unsteady process, the compositions of shock waves vary, making it difficult to achieve consistency. Researchers are also concerned about the flow unsteadiness, including its features, patterns, and occurrence in parameter space. Durna and Celik [12] suggested the existence of a critical aft wedge angle, specifically 47°, beyond which the interaction would become unsteady. While Kumar and De [15] disagreed with the proposed angle by [12], they also found that the combination of $(\theta_2 - \theta_1)$ and $L_1/L_2$ would determine whether the flow becomes unsteady, identifying three patterns numerically as vibration, oscillation, and pulsation [14]. Moreover, they obtained the configuration of these patterns in the parameter space through computation. Generally, studies such as [9, 12, 14] suggest that unsteadiness arises from the transmitted shock impinging on the aft wedge, leading to large adverse pressure gradients interacting with separation at the compression corner. This pushes the transmitted shock downstream, past the expansion corner; then, after the release of pressure, the shock moves backward, establishing periodic unsteadiness. Hence, a steady interaction with the transmitted shock residing on the aft wedge is unlikely. Most of the currently reported steady interactions, especially in 2D cases, involve transmitted shocks passing over the expansion corner [1, 14, 15].

Despite the aforementioned practices and achievements in M7_2, there exist uncertainties and deficiencies: (1) There is uncertainty regarding the accurate prediction and fidelity of certain parameters such as heat flux and shock wave structures. Although studies have shown some agreement between averaged heat transfer and experimental results at $327\mu s$, the temporal accuracy of the latter is definite attributed to the $1\mu s$ response time of the coaxial thermocouple. This raises doubts about the necessity of numerical averaging. Additionally, some researchers believe that the flow in experiments is still evolving due to the short test time, further complicating accuracy assessment. Hence, it is worthwhile to investigate a *steady* interaction with similar complexity as that in [2, 3] because the steady results are usually more certain and credible. (2) It is uncertain whether discrepancies would arise if a real gas model is used instead of a perfect gas model in cases of low enthalpy. Komives et al. [16] compared the results of perfect gas and thermal non-equilibrium gas models at 500 flowtimes and found nonconformity in surface heating downstream of the bow shock. This suggests that interaction nonlinearity may amplify the differences in physical models, which are usually absent ahead of the interaction. Although some comparisons in [9, 10] did not show significant differences with the non-equilibrium model, a single case of M7_2 may not be sufficient to fully understand the impact of real gas models. (3) Lastly, when considering variations in interaction with respect to the 30–55° wedge, the main concerns are those caused by geometric parameters, e.g., $L_1/L_2$ and/or $(\theta_2 - \theta_1)$; however, it is important to also consider variations in inflow conditions such as $Re$ and $Ma$. To the best of our knowledge, the only case that considered this was Tumukulu et al. [17], who used DSMC to study a case similar to M8_7 with the same $Ma$ but a lower $Re$ (by a factor of about eight). A steady interaction was obtained, where the transmitted shock resided on the second wedge before the expansion corner. Further investigation is warranted to understand the effects of comprehensive variations in inflow conditions on the *low enthalpy* counterpart.

In view of the aforementioned issues, referring to M7_2, the steady interaction of low enthalpy hypersonic flows over 2D 30–55° double wedge flows is studied herein using different gas models, including the perfect, equilibrium, and non-equilibrium gas models. This study addresses three key concerns: first, the analysis of the steady interaction pattern and its evolution with aerodynamic parameters such as $Re$, $Ma$, and $h_0$; second, the comparison of prediction differences resulting from different gas models; and third, the theoretical examination of interactions along with a comparison of predictions to computational results. This paper is organized as follows: Section 2 introduces the numerical schemes and physical models, followed by grid convergence studies to determine the appropriate grids for the main study in Section 3. Section 4 presents the numerical investigation of interaction pattern variations with decreasing $Re$ using the perfect, equilibrium, and non-equilibrium gas models. Section 5 focuses on a theoretical analysis of interactions and a comparison of the predictions with the computational results, while Section 6 explores further variations in interactions with $Ma$ and $h_0$. Finally, Section 7 provides the conclusions.

## 2 Governing equations, gas models, and numerical methods

### 2.1 Governing equations and gas models

As mentioned in the introduction, the hypersonic low enthalpy double wedge in this study experiences flows with low $Re$, approximately one fourth of or even smaller than that of M7_2. Therefore, it is reasonable to use laminar Navier–Stokes equations for simulations. Additionally, in terms of low enthalpy scenarios, the thermal non-equilibrium effect is ignored. For illustration, the governing equations for chemical non-equilibrium flows are presented below, from which the equations for a perfect gas can be derived. For a gas mixture comprising *ns* species, the equations are as follows:

$$\frac{\partial Q}{\partial t} + \frac{\partial E}{\partial x} + \frac{\partial F}{\partial y} + \frac{\partial G}{\partial z} - \left(\frac{\partial E_v}{\partial x} + \frac{\partial F_v}{\partial y} + \frac{\partial G_v}{\partial z}\right) = S \qquad (1)$$

where $Q = \begin{bmatrix} \rho_1 \\ \vdots \\ \rho_{ns} \\ \rho u \\ \rho v \\ \rho w \\ \rho E \end{bmatrix}$, $E = \begin{bmatrix} \rho_1 u \\ \vdots \\ \rho_{ns} u \\ \rho u^2 + p \\ \rho uv \\ \rho uw \\ \rho u h_0 \end{bmatrix}$, $F = \begin{bmatrix} \rho_1 v \\ \vdots \\ \rho_{ns} v \\ \rho vu \\ \rho v^2 + p \\ \rho vw \\ \rho v h_0 \end{bmatrix}$, $G = \begin{bmatrix} \rho_1 w \\ \vdots \\ \rho_{ns} w \\ \rho wu \\ \rho wv \\ \rho w^2 + p \\ \rho w h_0 \end{bmatrix}$, $S = \begin{bmatrix} \omega_1 \\ \vdots \\ \omega_{ns} \\ 0 \\ 0 \\ 0 \\ 0 \end{bmatrix}$, $E_v =$

$$\begin{bmatrix} q_{x1} \\ \vdots \\ q_{xns} \\ \tau_{xx} \\ \tau_{xy} \\ \tau_{xz} \\ u\tau_{xx} + v\tau_{xy} + w\tau_{xz} + \\ q_x + \rho \sum_{i=1}^{ns} D_i h_i \frac{\partial Y_i}{\partial x} \end{bmatrix}, \quad F_v = \begin{bmatrix} q_{y1} \\ \vdots \\ q_{yns} \\ \tau_{yx} \\ \tau_{yy} \\ \tau_{yz} \\ u\tau_{yx} + v\tau_{yy} + w\tau_{yz} + \\ q_y + \rho \sum_{i=1}^{ns} D_i h_i \frac{\partial Y_i}{\partial y} \end{bmatrix}, \quad G_v = \begin{bmatrix} q_{z1} \\ \vdots \\ q_{zns} \\ \tau_{zx} \\ \tau_{zy} \\ \tau_{zz} \\ u\tau_{zx} + v\tau_{zy} + w\tau_{zz} + \\ q_z + \rho \sum_{i=1}^{ns} D_i h_i \frac{\partial Y_i}{\partial z} \end{bmatrix}; \text{ and}$$

where $\rho_i$ is the density of the $i$-th species, $Y_i$ is its mass fraction ($\rho_i/\rho$), $\omega_i$ is the generation source term, and the mass diffusion is $q_{x_j i} = \rho D_i \partial Y_i / \partial x_j$, with $D_i$ as the diffusion coefficient and $h_i$ representing the specific enthalpy. Additionally, $h_0$ is the specific total enthalpy, which can be calculated as $\sum_{i=1}^{ns} Y_i h_i + (\sum_{j=1}^{3} u_j^2)/2$. The viscous stress is $\tau_{x_i x_j} = -\frac{2}{3}\mu \nabla \cdot \vec{V} + \mu(\frac{\partial u_i}{\partial x_j} + \frac{\partial u_j}{\partial x_i})$ with $\mu$ representing the viscous coefficient. Meanwhile, $p = \sum_{i=1}^{ns} p_i = \sum_{i=1}^{ns} \rho_i R_i T$ with $R_i$ representing the gas constant. Furthermore, $q_{x_j} = k\ \partial T / \partial x_j$ with $k$ representing the heat conductivity. In the case of the perfect gas and equilibrium gas models used in this study, the multispecies component is not considered, and the equations can be formally perceived by reducing $ns$ species to one. Consequently, $\rho_i$ becomes $\rho$, while $\omega_i$, $q_{x_j i}$ and $\partial Y_i / \partial x_j$ disappear. It should be noted that different gas models will result in distinct interpretations of transport coefficients and relationships between thermodynamic properties. The specific transport coefficients and relations employed in this study will be introduced subsequently.

(1) Perfect gas

In the simplest case, the viscous coefficient $\mu$ is determined by the Sutherland formula, while the heat conductivity is indirectly defined using $Pr$ or $k = \mu C_p / Pr$ with $Pr=0.72$ for air. The relationship between thermodynamic properties is described by the state equation $p = \rho RT$.

(2) Equilibrium gas

The equilibrium gas model is applicable to flows where the characteristic time scale of reactions is much smaller than that of the flow motion. Currently, the curve fitting method developed by Srinivasan et al. [18, 19] is being used, where thermodynamic and transport properties are determined from two independent variables using a Grabau type piecewise function defined in separate regions. In terms of thermodynamic properties, the temperature $T$ is considered as the dependent variable of pressure $p$ and density $\rho$ in this study, expressed as $\log_{10}(T/T_0) = f(X,Y)$, where $X = \log_{10}(\rho/\rho_0)$, $Y = \log_{10}(p/p_0) - X$, and where $p_0 = 1.1034 \times 10^5 N/m^2$, $\rho_0 = 1.292 kg/m^3$, and $T_0 = 273.15 K$. $f(X,Y)$ is a Grabau type function having the form: $f(x,y) = f_1(x,y) + \frac{f_2(x,y) - f_1(x,y)}{1 \pm \exp(k_0 + k_1 x + k_2 y + k_3 xy)}$, where "+" in the denominator takes effect in the case of an odd function and "-" acts under an even function. In $f(x,y)$, $f_1(x,y) = p_1 + p_2 x + p_3 y + p_4 xy + p_5 x^2 + p_6 y^2 + p_7 x^2 y + p_8 xy^2 + p_9 x^3 + p_{10} y^3$ and $f_2(x,y) = f_1(x,y) + p_{11} + p_{12} x + p_{13} y + p_{14} xy + p_{15} x^2 + p_{16} y^2 + p_{17} x^2 y + p_{18} xy^2 + p_{19} x^3 + p_{20} y^3$, where the coefficients $p_1 \sim p_{20}$ of air can be found in [18, 19]. The equivalent specific heat ratio, $\tilde{\gamma}$, is derived in a similar manner; using this, the sound speed can be obtained as $a = \sqrt{\tilde{\gamma} RT}$. In addition, the transport properties, $\mu$ and $k$, are acquired by curve fitting. For example, $\mu/\mu_0 = f(X,Y)$ where $X = T/1000K$ and $Y = \log_{10}(\rho/\rho_0)$, and where $\rho_0 = 1.243 kg/m^3$ and $\mu_0 = 17.486 \times 10^{-6} kg \cdot s/m$. More details can be found in [18, 19]. It should be noted that the abovementioned methods for determining thermodynamic and transport properties are considered to be valid for temperatures up to 25000 K and 15000 K, respectively, in the case of equilibrium flows.

(3) Chemical non-equilibrium gas

As the characteristic time scale of reactions is the same order as that of the flow motion, the chemical process with finite rates should be considered. Suppose the equations for *ns* species and *nr* reversible reactions are formulated as follows:

$$\sum_{i=1}^{ns} \alpha_{ji} A_i \rightleftharpoons \sum_{i=1}^{ns} \beta_{ji} A_i \quad (j = 1, \cdots, nr), \tag{2}$$

where $\alpha_{ji}$ and $\beta_{ji}$ are stoichiometric coefficients of reactants; those of colliders would not be considered. According to Eq. (2), the source generation of the *i*-th species in Eq. (1) is defined as follows: $\omega_i = M_i \sum_{j=1}^{nr} (d(\rho_i/M_i)/dt)_j$, where $M_i$ is the molecular weight and

$$\left(d(\frac{\rho_i}{M_i})/dt\right)_j = (\beta_{ji} - \alpha_{ji}) \left[ k_{f,j} (\prod_{m=1}^{ns} \left(\frac{\rho_m}{M_m}\right)^{\alpha_{jm}}) \left(\sum_{n=1}^{ns} \frac{\rho_n}{M_n} C_{nj}\right)^{L_j} - k_{b,j} (\prod_{m=1}^{ns} \left(\frac{\rho_m}{M_m}\right)^{\beta_{jm}}) \left(\sum_{n=1}^{ns} \frac{\rho_n}{M_n} C_{nj}\right)^{L_j} \right]. \tag{3}$$

In Eq. (3), $k_{f,j}$ and $k_{b,j}$ are the forward and backward reaction rates, respectively; $C_{nj}$ are three-body collision coefficients corresponding to each species in reaction *j*; and $L_j$ is the switch indicating the collision. According to the Arrhenius law, the reaction rates are given as

$$\begin{cases} k_{f,j} = A_{f,j} T_c^{B_{f,j}} e^{-\frac{E_{f,j}}{R_u T_c}} \\ k_{b,j} = A_{b,j} T_c^{B_{b,j}} e^{-\frac{E_{b,j}}{R_u T_c}} \end{cases}, \tag{4}$$

where $A_{f/b,j}$, $B_{f,j}$, and $E_{f,j}/R_u$ are reaction-related coefficients, and $T_c$ is the control temperature and actually $T$ here. In this study, an air model of 5 species and 6 reactions developed by Gupta et al. [20] is employed, with the reaction equations shown in Table 1. Here, $M_{1-3}$ denote the colliders, namely $M_1 = M_3 = \{O, N, O_2, N_2, NO\}$, $M_2 = \{O, O_2, N_2, NO\}$. Details on the coefficients can be found in [20]. Besides, the non-catalytic condition is employed at the wall considering the metal material of the double wedge.

Table 1. Reactions of the gas model with 5 species and 6 reactions [20]

| Index | Reaction equation |
|---|---|
| 1 | $O_2 + M_1 \rightleftharpoons 2O + M_1$ |
| 2 | $N_2 + M_2 \rightleftharpoons 2N + M_2$ |
| 3 | $N_2 + N \rightleftharpoons 2N + N$ |
| 4 | $NO + M_3 \rightleftharpoons N + O + M_3$ |
| 5 | $NO + O \rightleftharpoons O_2 + N$ |
| 6 | $N_2 + O \rightleftharpoons NO + N$ |

**2.2 Numerical methods**

In order to address the complex shock interactions in high-speed flows, it is critical to use numerical schemes that exhibit high resolution and robustness. Central schemes are commonly used for discretizing viscous terms, while the focus of research lies in developing schemes for computing the convection terms in Eq. (1). In this regard, a third-order scheme, WENO3-PRM$_{1,1}^2$, which was developed in [21], is applied in computations. To address the limitations of WENO3-JS in achieving optimal order in the presence of the first-order critical point ($CP_1$), the authors [21] conducted an analysis considering the occurrence of critical points within grid intervals. They then devised an improvement by introducing a new piecewise-rational polynomial mapping (PRM) from the perspective of WENO-M. WENO3-PRM$_{1,1}^2$ can achieve third-order accuracy at $CP_1$ regardless of its location in the stencil, demonstrating high resolution in resolving flow subtleties and exhibiting strong robustness in hypersonic simulations. The following is a brief overview of the scheme.

Considering the one-dimensional hyperbolic conservation law $u_t + f(u)_x = 0$ and supposing $\partial f(u)/\partial u > 0$, the conservative approximation of $f(u)_x$ at $x_j$ can be expressed as

$$\frac{\partial f(u)}{\partial x_j} \approx \frac{(\hat{f}_{j+1/2} - \hat{f}_{j-1/2})}{\Delta x} \tag{5}$$

where $\hat{f}_{j+1/2}$ denotes some numerical flux. Assuming that $r$ is the number of sub-stencils and the grid number of each stencil, WENO-JS schemes with an order of $r$ can be formulated as $\hat{f}_{j+1/2} = \sum_{k=0}^{r-1} \omega_k q_k^r$ with $q_k^r = \sum_{l=0}^{r-1} a_{kl}^r f(u_{j-r+k+l+1})$, where $q_k^r$ is the candidate scheme with coefficients $a_{kl}^r$ and $\omega_k$ is the normalized nonlinear weight corresponding to the linear counterpart $d_k^r$. $\omega_k$ is typically derived from the non-normalized weight $\alpha_k$. For the WENO3-JS scheme where $r = 2$, $\alpha_k = d_k^r / (\varepsilon + IS_k^{(r)})^2$ and $IS_k^{(r)} = \sum_{m=0}^{r-2} c_m^r [\sum_{l=0}^{r-1} b_{kml}^r f(u_{j-r+k+l+1})]^2$. Here, $\varepsilon$ is a small quantity included to prevent the denominator from becoming zero, and the values of $a_{kl}^r$, $b_{kml}^r$, $c_m^r$, and $d_k^r$ as well as other coefficients can be found in [21]. As reviewed in [21], to achieve the optimal order of WENO-JS at $CP_l$, Henrick et al [22]. introduced a mapping function for the nonlinear weight according to the following principles: (a) For $\omega \in [0, 1]$, consider some mapping function $g_k(\omega)$ with respect to the linear weight $d_k$ satisfying $g_k(\omega) = \omega$ when $\omega = \{0, d_k, 1\}$. If there exists a constant $n$ such that $g_k^{(i)}(\omega) = 0$ and $g_k^{(n)}(\omega) \neq 0$ when $1 \leq i < n$, then it holds in the neighborhood of $d_k$ that $g_k(\omega) = d_k + (1/n!)g_k^{(n)}(\omega - d_k)^n + \cdots$; (b) Considering $g_k(\omega)$ as a new non-normalized weight $\alpha_k'$ about $g_k(\omega)$, then $\alpha_k' = d_k + O(\Delta x^{n \times m})$, providing $\omega - d_k = O(\Delta x^m)$ with $m \geq 1$, and a new nonlinear weight $\omega_k'$ can be obtained after normalization, with $\omega_k' = d_k + O(\Delta x^{n \times m})$; (c) Hence, after mapping, the original order of $(\omega - d_k)$ is improved from $n$ to $m \times n$. It can be conceivable the mapping function is critical in achieving high resolution while preserving numerical robustness. Considering this, the so-called WENO-PRM schemes are developed in [21] through the use of PRM. This new mapping has the following form:

$$\begin{cases} PRM_{n,m;m_1;c_1,c_2}^{L,n+1} = d_k + \frac{(\omega-d_k)^{n+1}}{(\omega-d_k)^n + (-1)^{1+n}c_2^L(\omega-d_k)\omega^{m_1} + (-1)c_1^L\omega^{m+1}} & \text{when } \omega < d_k \\ PRM_{n,m;m_1;c_1,c_2}^{R,n+1} = d_k + \frac{(\omega-d_k)^{n+1}}{(\omega-d_k)^n + c_2^R(\omega-d_k)(1-\omega)^{m_1} + c_1^R(1-\omega)^{m+1}} & \text{when } \omega \geq d_k \end{cases} \tag{6}$$

where superscript $L$ indicates that the function is for $\omega < d_k$, and the superscript $R$ indicates that it is for $\omega \geq d_k$. Additionally, $n$ corresponds to the constant $n$ in relation to the previous $g_k(\omega)$. By adjusting the parameters $c_1, c_2$, and $m_1$ in Eq. (6), the mapping function exhibits good flatness at $d_k$ and an optimized characteristic when $\omega$ approaches the endpoints $\{0,1\}$. Therefore, the mapped WENO schemes achieve high accuracy, resolution, and robustness. In WENO3-PRM$_{1,1}^2$, where $n = 1$ and $m = 1$ in Eq. (6), the other parameters are as follows: for $d_k=1/3$, $(c_1, c_2, m_1) = (1, 7 \times 10^7, 5)$ when $0 < \omega < d_k$; otherwise, $(c_1, c_2, m_1) = (1, 3 \times 10^6, 5)$; for $d_k = 2/3$, $(c_1, c_2, m_1) = (1, 1 \times 10^5, 4)$ when $0 < \omega < d_k$; otherwise, $(c_1, c_2, m_1) = (1, 3 \times 10^6, 4)$. In addition, to satisfy $\omega - d_k = O(\Delta x^m)$ with $m \geq 1$, $IS_k^{(3)}$ should be used [21].

Additionally, given that flow steadiness is the main focus of this study, the temporal discretization uses the canonical LU-SGS method with a local time step to efficiently determine flow steadiness.

## 2.3 Validation tests

Although the validations and applications of WENO3-PRM$_{1,1}^{2}$ have been provided in [21, 23], its performance on the double cone flow at $Ma = 9.59$ (Run28) is shown here for demonstration. Additionally, a hypersonic equilibrium flow over a sharp cone and a chemical non-equilibrium flow around a sphere are tested to verify the implementation of gas models.

(1) Double cone flow at $Ma = 9.59$

This case often serves as a test to demonstrate the capability of the scheme to resolve shock interactions. The inflow conditions are: $Ma = 9.59$, $Re = 1.44 \times 10^5/m$, $T_\infty = 42.6$ K, and $T_w = 293.3$ K. A typical $256 \times 128$ grid is chosen for this test. For reference, the result of WENO3-JS is shown.

The numerical schlieren of WENO3-PRM$_{1,1}^{2}$ is shown in Fig. 1, which qualitatively displays the shock interaction structure consisting of the leading-edge shock, the separation shock, and the flare shock. Moreover, the Edney type IV interaction and corresponding supersonic jet are indicated. In Fig. 2, the heat flux predictions by WENO3-PRM$_{1,1}^{2}$ and WENO3-JS are compared to the experimental results. It is evident that WENO3-PRM$_{1,1}^{2}$ reasonably predicted the separation scale and peak heat transfer, while WENO3-JS resulted in a smaller separation with a delayed start and early termination. This demonstrates the capability of WENO3-PRM$_{1,1}^{2}$ to resolve interactions effectively.

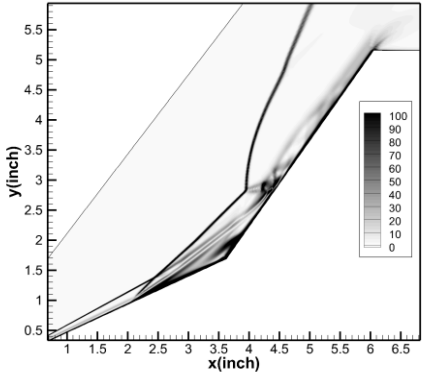
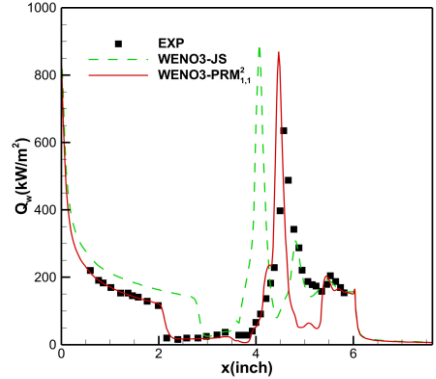

Fig. 1 Density gradient contours of double cone in case Run28 by WENO3-PRM$_{1,1}^{2}$

Fig. 2 Heat flux predictions of double cone in case Run28 compared with experimental results [24]

(2) Hypersonic cone flow at $Ma = 25.3$ and $0°$ angle of attack

This case involves an air flow around a cone with a half cone angle of $10°$, under the following air inflow conditions: $Ma = 25.3$, $Re = 1.29 \times 10^5$, $T_\infty = 252.6$ K, and $T_w = 1200$ K. In Fig. 3, the distribution of $T/T_\infty$ is derived from computations and compared with the results of [25]. In Fig. 4, the heat transfer coefficient is shown along the $x$ direction and also compared with the results of [25]. The comparisons indicate that the current predictions are in good agreement with the reference, validating the implementation of the equilibrium gas model.

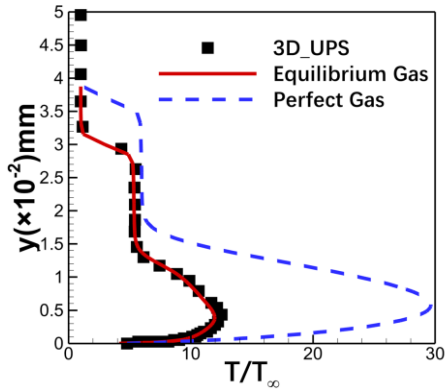 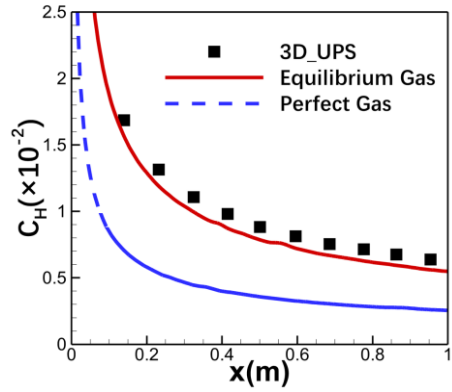

Fig. 3 Distribution of $T/T_\infty$ along $y$ direction at $x = 1m$ compared with the results of [24]

Fig. 4 Distribution of $Q/(\frac{1}{2}\rho_\infty u_\infty^3)$ along $x$ direction compared with the results of [24]

(3) Hypersonic flow around half sphere at $Ma = 15.3$

This case is usually used to validate the implementation of chemical non-equilibrium. The air inflow conditions are as follows: $Ma = 15.3$, $T_\infty = 293$ K, $T_w = 2000$ K, $Re(/m) = 2.25 \times 10^6/m$. In Fig. 5, the mass fractions of species along the stagnation line are shown and compared with the results produced by HYFLOW [26], showing a reasonable agreement. Similarly, in Fig. 6, the temperature distribution along the same line is drawn and compared with the results from HYFLOW [26] as well. In conclusion, the chemical non-equilibrium implementation appears to be validated.

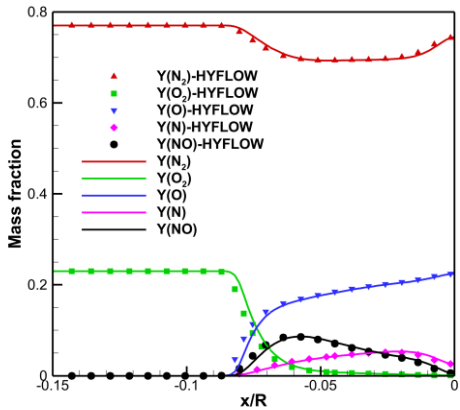 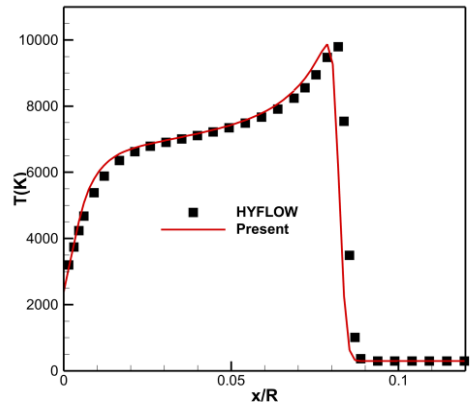

Fig. 5 Distributions of mass fractions of species along the stagnation line compared with the results of HYFLOW [26]

Fig. 6 Temperature distribution along the stagnation line compared with the results of HYFLOW [26]

## 3  Grid convergence studies at $Ma = 7$, $h_0 = 2.1 MJ/kg$, and $Re = 2.5 \times 10^5/m$

Before conducting the main study, it is essential to conduct a grid convergence study to determine the appropriate grids. Considering the objectives of this study, investigations using three gas models were carried out separately. Moreover, to account for the possibility of a transmitted shock impinging at a far downstream location, the horizontal plate after the expansion corner was extended, and the geometry of the double wedge is characterized as follows: $(\theta_1, \theta_2) = (30°, 55°)$ for two wedge angles; $L_1 = 50.8$ mm, $L_2 = 25.4$ mm, and $L_3 = 21.64$ mm for the lengths of the three sections. As highlighted in the introduction, this study is primarily focused on steady interactions rather than unsteady flows. Therefore, conditions with the same $Ma = 7$, $h_0 = 2.1 MJ/kg$ but smaller $Re$ are considered, specifically, $Re = 2.5 \times 10^5/m$ for the grid study, or even smaller for the main study. The remaining inflow conditions are specified as follows: $p_\infty = 90.25 Pa$, $T_\infty = 191$ K, and $T_w = 298$ K. Following preliminary trials, the following grid schemes are planned: coarse grids as $541 \times 256$, medium grids as $812 \times 382$, and fine grids as $1083 \times 512$. The first normal grid interval is set at 0.001 mm.

The computations are carried out until steady solutions are achieved, and the heat transfer coefficients are then derived as shown in Fig. 7. For brevity, the notation "Chemical non-equilibrium" is abbreviated as "Non-equilibrium" hereon. From the figure, it can be observed that for an equilibrium gas, the computations on medium and fine grids produce consistent results that differ from those on coarse grids. However, for other gas models, all three grids yield almost consistent results. Hence, medium grids can be considered to achieve grid convergence for the three gas models. Although convergence can also be achieved with coarse grids, we have chosen to use medium grids for the main study due to the anticipated increase in interaction complexity. In Fig. 7(d), the results of different gas models on medium grids are depicted, with similar distributions indicated. The equilibrium gas model shows a slightly smaller separation.

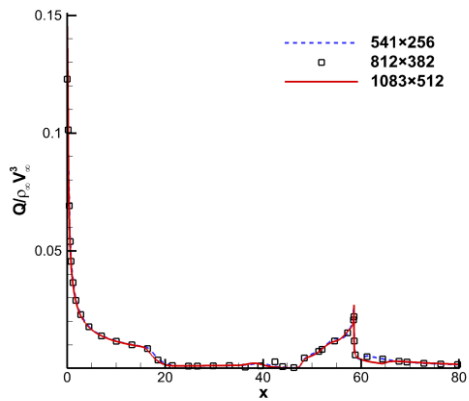
(a) Perfect gas

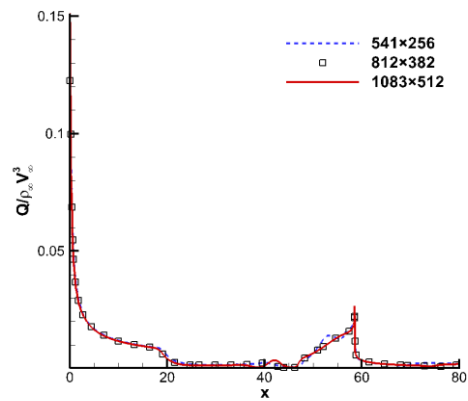
(b) Equilibrium gas

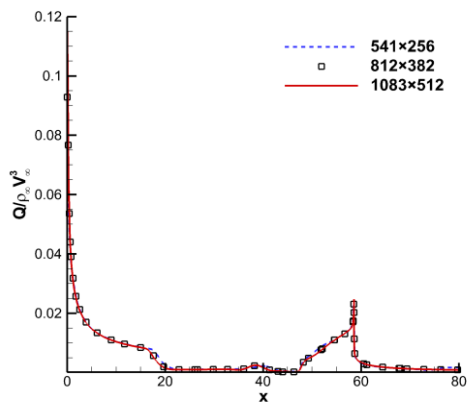
(c) Non-equilibrium gas

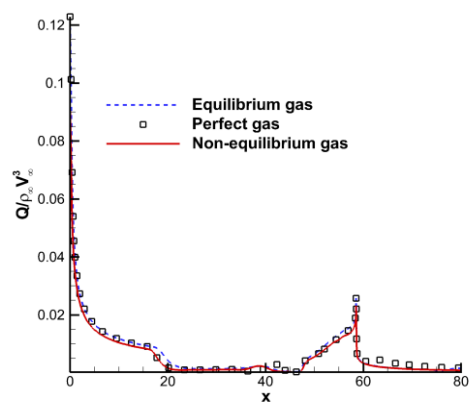
(d) Results of three gas models at $812 \times 382$

Fig. 7 Heat transfer coefficients of different gas models on three different types of grids and a

comparison on medium grids

One may wonder how the flow structures would appear with regard to the distributions mentioned above. To this end, the numerical schlieren of three gas models on medium grids are shown in Fig. 8, with the sonic lines drawn. For reference, the quasi-steady interaction of M7_2 from the courtesy of [15] is shown in Fig. 8(d). Moreover, the shock outlines of perfect gas are derived and overlapped with the other results. From Fig. 8(a)–(c), although the three gas models yield qualitatively approximate configurations, the shapes in Fig. 8(b)–(c) differ from that in Fig. 8(a), especially regarding the bow shock and transmitted shock. Referring back to Fig. 7(d), it is indicated that the consistency of heat transfer distributions does not necessarily imply a similar establishment of the shock structure. In terms of shock interaction, it appears that the three gas models exhibit similar patterns, namely, a local subsonic flow situated between two supersonic neighbors, resembling an Edney type IV-like interaction to some extent. However, the subsonic region after the bow shock causes the interaction to deviate from the standard pattern. From Fig. 8(b)–(c), it can be observed that the local subsonic regions are generated after the intersection of the compression/shock wave and transmitted shock, which also differs from the jet-like appearance in canonical Edney type IV interactions. Regarding the shock profiles, it is evident that the gas models have little effect on the formation of the leading-edge shock. Furthermore, the separation shock of the non-equilibrium model intersects the leading-edge shock at the same location as that of the perfect gas model, whereas the intersection is slightly delayed in the equilibrium gas model. The separation shocks in Fig. 8(b)–(c) have a relatively smaller slope than in the perfect gas model, with the triple points of the former occurring more downstream, and their bow shocks closer to the wedge. Fig. 8(d) reveals that the larger $Re$ in perfect gas flows results in a larger separation, leading to the early intersection of the leading-edge and separation shocks, as well as a delayed occurrence of the triple point.

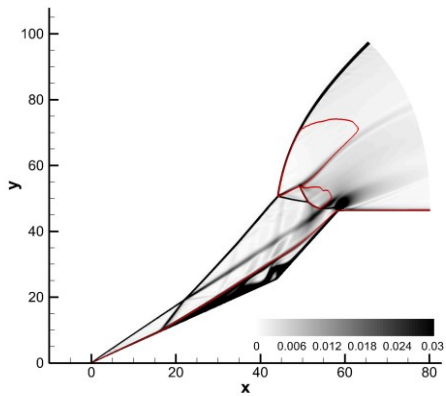
(a) Perfect gas

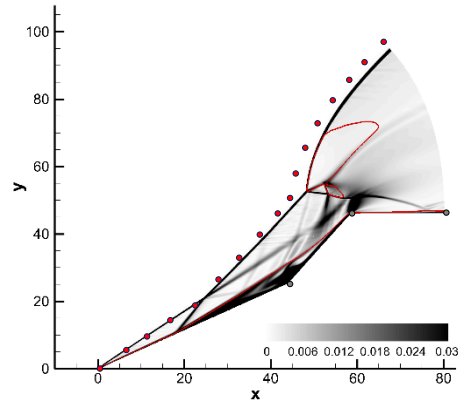
(b) Equilibrium gas

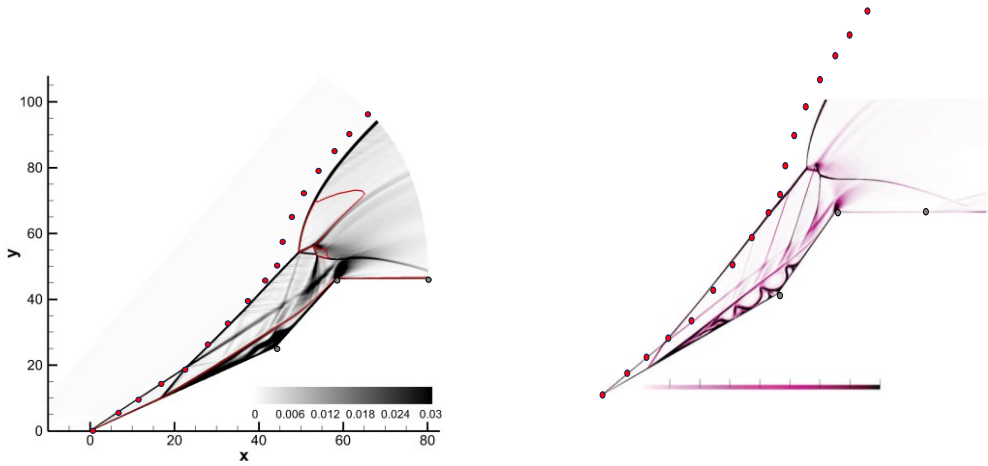

(c) Non-equilibrium gas      (d) Perfect gas of M7_2 from [15]

Fig. 8 Numerical schlieren by three gas models with the sonic lines (red) indicated and that of perfect gas of M7_2 courtesy of [15], where the shock profile (represented by circles) of the perfect gas model are superimposed upon those of the real gas model

In summary, the grid convergence of three gas models is investigated, and medium grids are chosen for the main study. It is noteworthy that the shock configurations of the three gas models differ from each other to some extent.

## 4 Steady interactions and their evolution with low $Re$ at $Ma = 7$ and $h_0 = 2.11 MJ/kg$

As discussed in the introduction, this study focuses on investigating steady interactions having similar complexity to those in [2, 3]. To align with the experiment as closely as possible, the same $Ma$ and $h_0$ are employed, whereas $Re$ are lower than their counterpart. For completeness, the other parameters are as follows: $T_\infty = 191$ K, $T_w = 298$ K, $p_\infty = 14.4399 pa$. Using the methods outlined in Section 3, computations using three different gas models will be carried out. First, the upper limits of $Re$ where the flows remain steady need to be defined numerically. In this regard, a series of $Re$ are chosen with the lower limit of $2.0 \times 10^4/m$ and an incremental interval of $0.5 \times 10^4/m$. The lower limit is selected to ensure that the flow structure will exhibit inconspicuous changes if a smaller $Re$ is employed. As a result, a diagram indicating whether the flow will be steady for the three gas models can be obtained, as shown in Fig. 9. Based on this, the upper limits of $Re$ are found to be $9.5, 5.5,$ and $4 \times 10^4/m$, respectively, for the three gas models. It is important to highlight that each gas model exhibits a different upper limit, even under conditions of low enthalpy. This underscores the nonlinearity of the interaction, highlighting how small differences among gas models can be amplified.

Numerical simulations show that as $Re$ increases before the flow becomes unsteady, the transmitted shock moves towards the expansion corner. Therefore, it is important to track the trajectory of the impingement point of the shock with respect to the corner. This process is illustrated in Fig. 10, where a coordinate transformation is made by setting the $x'$-axis along the aft wedge with the origin at the expansion corner. Additionally, Fig. 11 shows the distances of the impingement point from the corner for three models, as well as the angle between the line connecting the point to the corner and the aft wedge surface. The impingement point is defined as the location where the transmitted shock reflects above the boundary layer. The figures indicate that as the $Re$ increases, all impingement points move towards a specific region, namely [54.4~55.3 mm, 44~44.6 mm] or the equivalent de as shown in Fig. 11. Further increases of the $Re$ will result in a unsteady flow, with the transmitted shock moving back and forth around the expansion corner. Reflecting on the steady case discussed in Section 3, it can be inferred that within a certain range of $Re$, higher than the limits depicted in Fig. 9 but lower than $Re = 2.5 \times 10^5/m$, the flow with the transmitted shock inclining to impinge the expansion corner is unstable, which evolves into unsteady interactions.

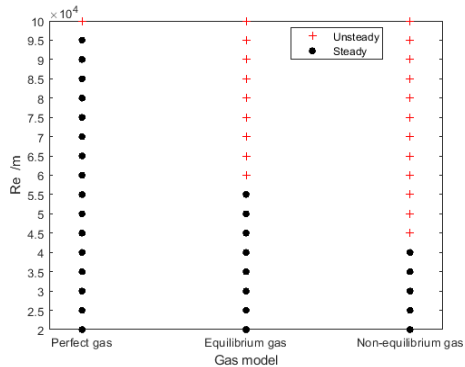

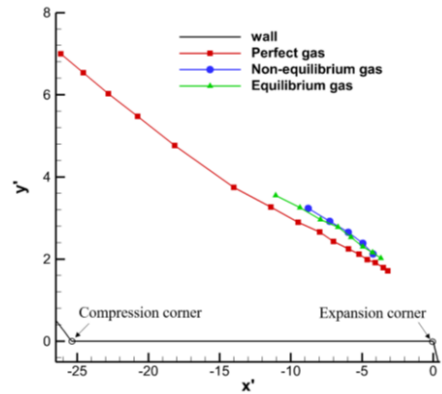

Fig. 9 Flow steadiness vs unsteadiness of three gas models at $Ma = 7$ and $h_0 = 2.1 MJ/kg$: "•" – steady; "+" - unsteady

Fig. 10 Coordinate trajectories of impingement point of transmitted shock of the three gas models

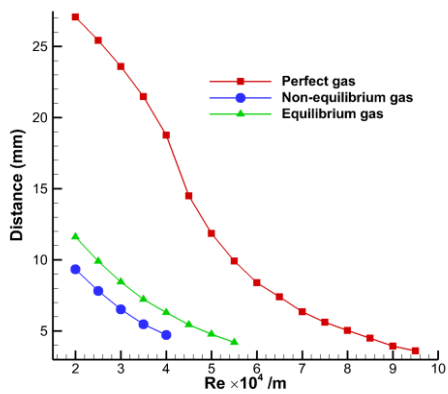

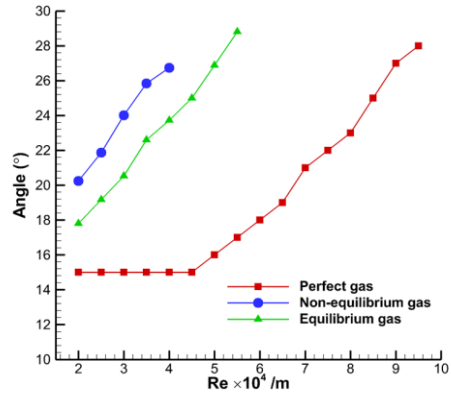

(a) Distance of the point to the expansion corner

(b) Angle between line of the point to the expansion corner and aft wedge surface

Fig. 11 Variation in geometric properties of impingement point with $Re$ of three gas models

The subsequent discussions will focus on the details of the interactions.

(1) Interaction pattern and flow structures

Considering the evolving nature of the steady flow with $Re$, the discussions will be organized as follows: first, an analysis of flows at the upper limits of $Re$ for the three gas models will be conducted. Following this, comparative studies will be carried out on flows with the same three $Re$ to illustrate interaction evolutions. Lastly, the geometric characteristics of interaction and their variations with $Re$ will be discussed.

(1.1) Interactions at the upper limit of the $Re$ for the three gas models

In Fig. 12, the pressure contours of perfect and equilibrium gas models are shown, with the separation zone indicated by streamlines. The vortices are magnified in Fig. 13, with the numerical schlieren as the background, and the sonic line depicted. Similarly, the pressure contours of the non-equilibrium gas model are shown in Figs. 18 and 19 in the column where $Re = 4 \times 10^4/m$. In the figures, two slip lines are displayed by streamlines originating from the intersection of the leading-edge shock (LS) and the separation shock (SS), as well as the intersection of the combination of LS and SS (referred to as CS) and the bow shock (BS). Moreover, a magnified window is provided in Fig. 12(a) to clarify details around the expansion corner. Overall, the following characteristics are observed: the main framework of the shock waves consists of LS, SS, CS, BS, and the transmitted shock (TS). It is evident that SS is caused by the separation at the compression corner, while TS impinges the aft wedge at a position closest to the expansion corner to maintain a steady interaction. As indicated in Figs. 10 and 11, the details of impingement of the three gas models are different. To further illustrate these differences, the shock outline of the perfect gas model is extracted using circles and superimposed upon the results of the real gas models, as demonstrated in Fig. 12(b) and 18(c). The figures indicate that the perfect gas model yields an outline with the largest size, but it is relatively close to that of the non-equilibrium gas model. To further analyze the correlation between the shock interaction and the underlying vortex, corresponding details are checked in Fig. 13 and Fig. 19(c). The figures reveal that the three gas models yield different locations of separation onset, namely, the earliest being by the perfect gas model, followed by the equilibrium gas model, and lastly by the non-equilibrium gas model, as also indicated in Fig. 25. Moreover, a secondary separation is indicated by the perfect gas model. Despite the differences, the separations approximately end at a similar location near [56.64~57.14 mm, 43.46~44.19 mm].

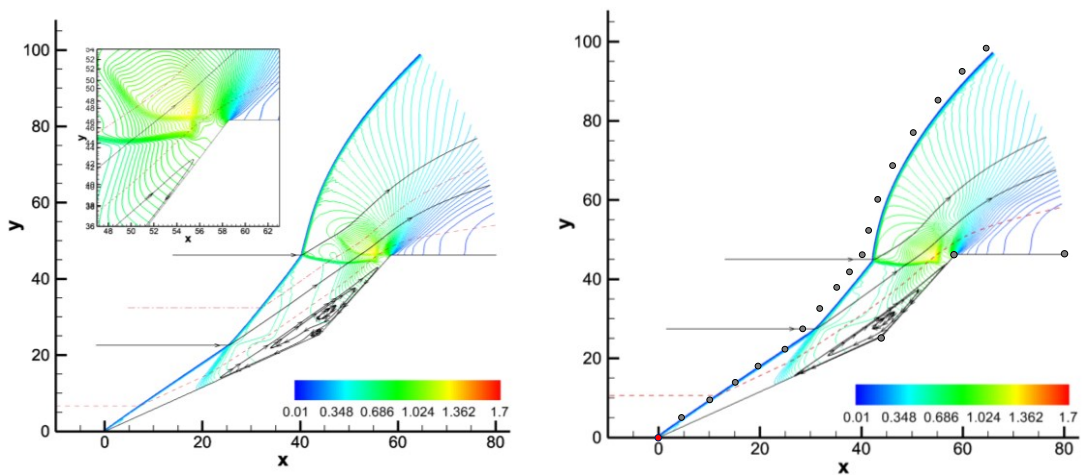

(a) Perfect gas at $Re = 9.5 \times 10^4/m$     (b) Equilibrium gas at $Re = 5.5 \times 10^4/m$

Fig. 12 Pressure contours of the perfect and equilibrium gas models at the upper limits of $Re$ of steady flows, where the shock profiles in dark circles of the perfect gas model are superimposed upon results of the equilibrium gas model, vortices and slip line tracks are shown by streamlines, and red dashed lines are chosen to display pressure and $Ma$ distributions

One interesting observation in Fig. 13 and Fig. 19(c) is the reflection behavior of the transmitted shock. The figures illustrate that the reflection occurs at the sonic line, which is located slightly ahead of the reattachment point. As shown in Figs. 12–13 and 18(c)–19(c), this reflection will evolve into an expansion wave and refract through the lower slip line. In the case of the perfect gas model, a quasi-normal shock (QNS) forms between the lower and upper slip lines, which was barely reported before. A subsonic region can be observed with a ceiling connected to that generated by BS. It can be anticipated that the QNS will create a pressure gradient that penetrates the lower slip line and evolves into a refracted wave (refer to the magnified window in Fig. 12(a)). This refracted wave then intersects with the TS reflection and dissipates in the expansion fan at the aft wedge corner. A theoretical analysis and flow model regarding this phenomenon will be presented in Section 5.

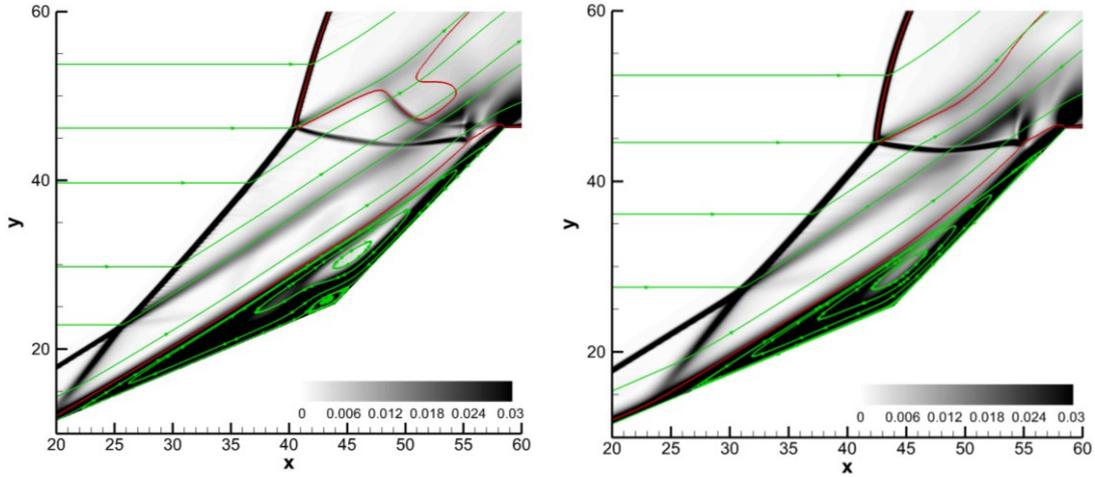

(a) Perfect gas at $Re = 9.5/m$　　　　(b) Equilibrium gas at $Re = 5.5/m$

Fig. 13 Vortex structures of perfect and equilibrium models at the upper limit of $Re$ of steady flows with sonic lines shown in red and with the background contours of $|\nabla \rho|$

In addition to the qualitative discussion above, a quantitative analysis is carried out by choosing two types of streamlines represented by dashed lines to demonstrate variable distributions, i.e., the one across TS and its corresponding reflections for three models, and the other across QNS of the perfect gas model in the middle. For illustration, these lines are shown in Fig. 13 and the first column of Fig. 18(c). It is important to note that the former streamlines may not align perfectly due to variations in TS and its reflections across the three cases. The distributions of pressure and $Ma$ are chosen and shown in Fig. 14. Taking the perfect gas model as an example, the pressure distribution reveals the following: (a) Two distinct discontinuities, identified as LS and SS, are clearly present, e.g., those at $x \approx 7$ and $22\ mm$, respectively, in the perfect gas case, with the latter indicating a weaker strength. (b) A plateau with oscillations appears after SS. Upon closer inspection, it is observed that the drop therein corresponds to the incidence of a compression wave emerging from the intersection of LS and SS (refer to Figs. 12(a) and 13(a)). The plateau terminates with compression waves characterized by a slanted gradient. (c) A sharp rise is observed upon encountering TS, with its location consistent with the contours in Figs. 12 and 13. The overall drop in pressure after the peak for the three models suggests that the reflection of TS is an expansion wave. Another small increase within the decline phase of the perfect gas case is attributed to a refracted oblique shock, as shown in Fig. 12(a). The results of $Ma$ convey a similar pattern, as shown in Fig. 13(b), with characteristic positions coinciding with those of pressure. For clarity, the positions of shocks in the perfect gas model are marked in Fig. 14.

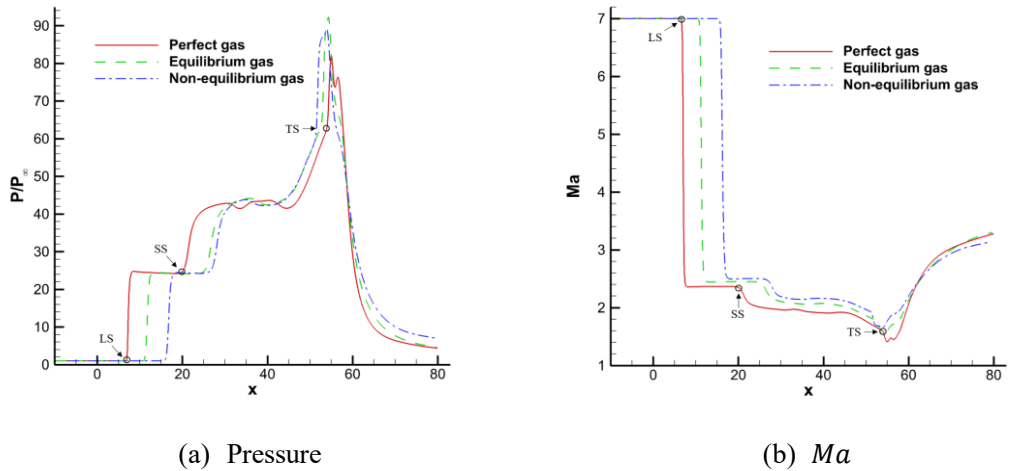

(a) Pressure  (b) $Ma$

Fig. 14 Variable distributions on chosen streamlines across the transmitted shock and its reflection of the three gas models

The variable distributions along the streamline passing the intensive part of QNS, represented by the upper dashed line in Fig. 12(a), are also shown in Fig. 15. The positions of SS, TS, and QNS are also marked. In the magnified window in Fig. 15(b), the subsonic region is indicated, with a length of approximately 4 mm. Additionally, $Ma$ having a value slightly less than 1 in that region suggests a weak solution of the shock relation.

In summary, this study analyzes the interaction characteristics of three gas models at the upper limits of $Re$, where the corresponding solutions remain steady and the impingements of transmitted shock waves is closest to the expansion corner. Different limits of $Re$ imply that diverse gas models result in different interactions even under low enthalpy conditions. To illustrate this point, Fig. 16 shows the contours of $\gamma$, revealing that the thermodynamic properties of the real gas solutions somewhat deviate from those of the perfect gas by 1.4.

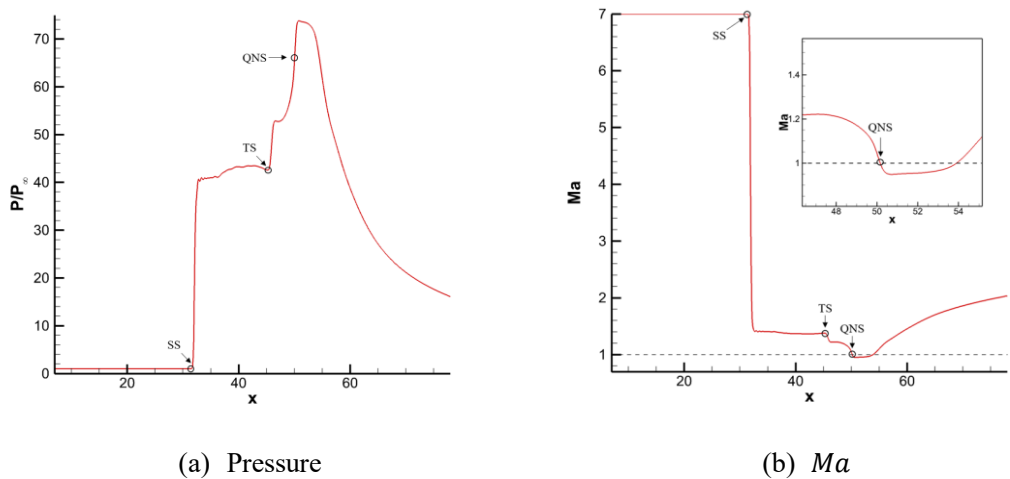

(a) Pressure  (b) $Ma$

Fig. 15 Variable distributions on streamline across the middle of quasi-normal shock of the perfect gas model

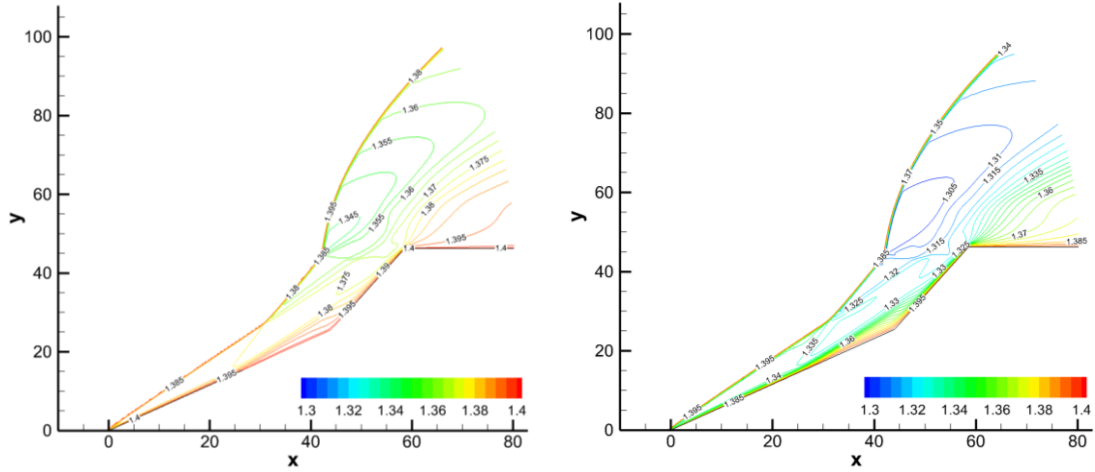

(a) Equilibrium gas at $Re = 5.5 \times 10^4/m$    (b) Non-equilibrium gas at $Re = 4 \times 10^4/m$

Fig. 16 Contours of $\gamma$ of real gas models at upper limit $Re$ under $Ma = 7$ and $h_0 = 2.1MJ/kg$

(1.2) Variations in interactions with $Re = 4, 3,$ and $2 \times 10^4/m$ of the three gas models

As indicated above, $Re$ plays an important role in achieving flow steadiness. It is natural to wonder how the steadiness will evolve with the decrease of $Re$, such as the upstream movement of the triple point and transmitted shock, the shrinkage of separation, etc. Considering the disparity of the upper limit of $Re$ between the perfect gas and real gas models, different performances would be expected. To illustrate this point, the typical characteristics of the perfect gas model are shown in Fig. 17, where $Re = 7.5, 5,$ and $2.5 \times 10^4/m$. For comparison, the shock outline at $Re = 7.5 \times 10^4/m$ is extracted and superimposed upon the others in circles. The following qualitative features are observed: (a) Enlargement and upstream movement of BS, resulting in the shortening of CS, enlargement of the subsonic region after BS, and closer interaction with the wedge; (b) Rapid shrinking of the previously existing subsonic region between two slip lines until it disappears; (c) Emergence of unusual interaction where the transmitted shock impinges and reflects over the separation, at least at $Re = 2.5 \times 10^4/m$; (d) Minor changes in the location of LS. Given the above situation, we further analyze the performances of the real gas models. Due to the smaller upper limit of $Re$ in these models, $Re$ in the subsequent studies are uniformly chosen as: 4, 3, and $2 \times 10^4/m$, with the first $Re$ as the upper limit of the non-equilibrium gas model.

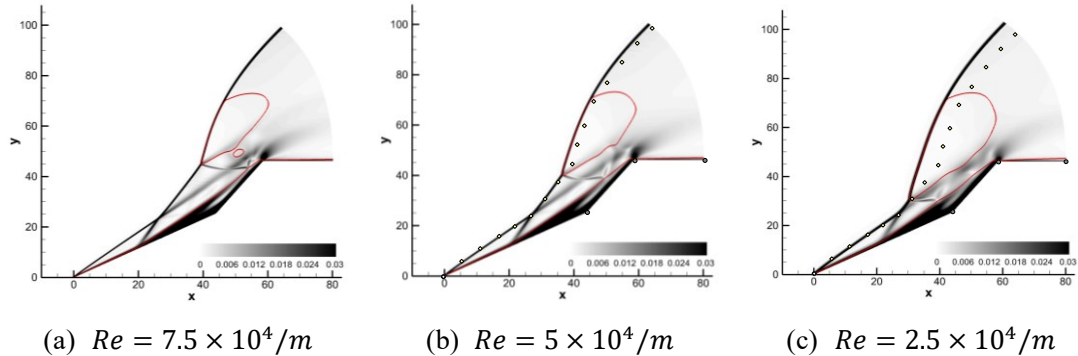

(a) $Re = 7.5 \times 10^4/m$    (b) $Re = 5 \times 10^4/m$    (c) $Re = 2.5 \times 10^4/m$

Fig. 17 Variations in density-gradient contour with $Re$ of the perfect gas model with sonic lines shown in red at $Ma = 7$ and $h_0 = 2.1MJ/kg$ and the shock outline at $Re = 7.5 \times 10^4/m$ in circles superimposing on the remaining two results

In Fig. 18, the pressure contours of three models at three $Re$ are displayed to illustrate the shock system, with vortices and slip lines shown by streamlines. To compare shock waves, the shock outlines of the reference perfect gas model are extracted and superimposed on the related results. The vortices and flow separations around the compression corner are visualized in Fig. 19 in accordance with the shock wave. The shock system exhibits the following features: (a) Decreasing the $Re$ results in similar behavior in the shock outline of the perfect gas model, albeit to a lesser extent; however, the equilibrium and non-equilibrium gas models show relatively fewer changes, except for some reduction of CS and upstream movement of the triple point along the wedge surface. The shock outlines of real gas models at the same $Re$ overall resemble each other but differ from those of the perfect gas model, such as a less changed position of the triple point and TS and shock interactions closer to the wedge surface. (b) While the TS of all models move upstream along with the triple points as $Re$ decreases, the movements of the real gas model lag behind notably. A particular pattern emerges, where TS impingement occurs above the separation and slightly ahead of the reattachment point, resulting in a reflection in the form of an expansion wave, as further explained later. (c) Comparatively, noticeable changes occur in the perfect gas model as the $Re$ decreases; at the smallest $Re$, LS, SS, and BS further approach each other and make CS implicit. Moreover, a special interaction involving TS takes place, which will be discussed in the following section. Corresponding vortex structures are shown in Fig. 19 with the numerical schlieren as the background and the sonic line illustrated. In general, all vortices decrease in size as $Re$ decreases, with those of the perfect gas model appearing comparatively larger. The impingement and subsequent reflection of TS are detailed in the amplified visualization, which is consistent with the preceding discussion from pressure contours and will not be reiterated for brevity.

The interaction of the perfect gas model exhibits an unusual behavior, which can be seen in the case where $Re = 4 \times 10^4 /m$. Several key features emerge from this interaction: (a) TS impinges upon the shear layer over the separation slightly after the vortex center, leading to a reflection that passes through the lower slip line and dissipates between the two slip lines. (b) Downstream of the reflection, compression waves form. Some of these waves pass through the lower slip line, while others reflect and become oblique shock wave. This reflection, abbreviated as RS, hits the position near the expansion corner and dissipates within the expansion waves. A detailed sketch will be presented later in Fig. 33 in this paper. (c) As $Re$ decreases, a similar pattern is maintained, but some variations arise. These include the upstream movement of TS impingement and the attenuation of the reflections of TS and RS.

Referring to the interaction characteristics at the upper limit of $Re$ discussed in "(1.1)" above, it is important to highlight a significant difference observed between various gas models. Specifically, the impingement of TS in the case of a perfect gas model shifts from the reattachment region to the bulk of the shear layer above the separation vortex, resulting in an unusual reflection. On the other hand, real gas models exhibit consistent impingement near the reattachment point despite the shared upstream movement of TS and reduction in separation.

| $Re = 4 \times 10^4 /m$ | $Re = 3 \times 10^4 /m$ | $Re = 2 \times 10^4 /m$ |
| --- | --- | --- |

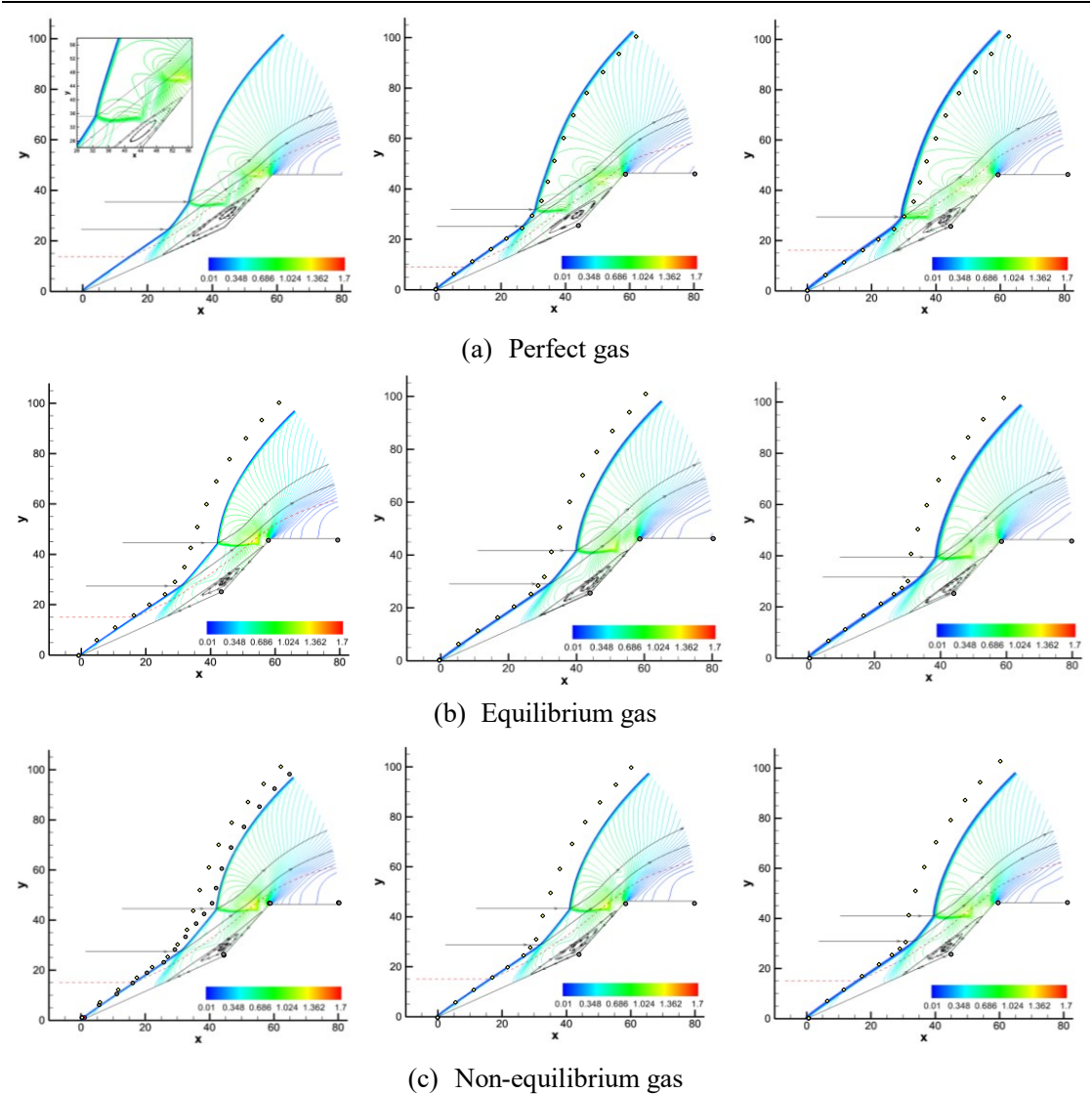

(a) Perfect gas

(b) Equilibrium gas

(c) Non-equilibrium gas

Fig. 18 Variations in pressure contour with $Re$ using three gas models at $Ma = 7$ and $h_0 = 2.1 MJ/kg$ where the shock outline of the perfect gas model at $Re = 4 \times 10^4/m$ is superimposed on cases of the remaining two $Re$ in diamonds, that of the perfect gas model of each $Re$ is superimposed on the results of equilibrium and non-equilibrium gas models with the same $Re$ in diamonds, and that of the perfect gas model at $Re = 9.5 \times 10^4/m$ is superimposed on the results of the non-equilibrium gas models at $Re = 4 \times 10^4/m$ in dark circles; vortices and slip lines are shown by streamlines, and red dashed lines are chosen to display pressure and $Ma$ distributions

| $Re = 4 \times 10^4/m$ | $Re = 3 \times 10^4/m$ | $Re = 2 \times 10^4/m$ |

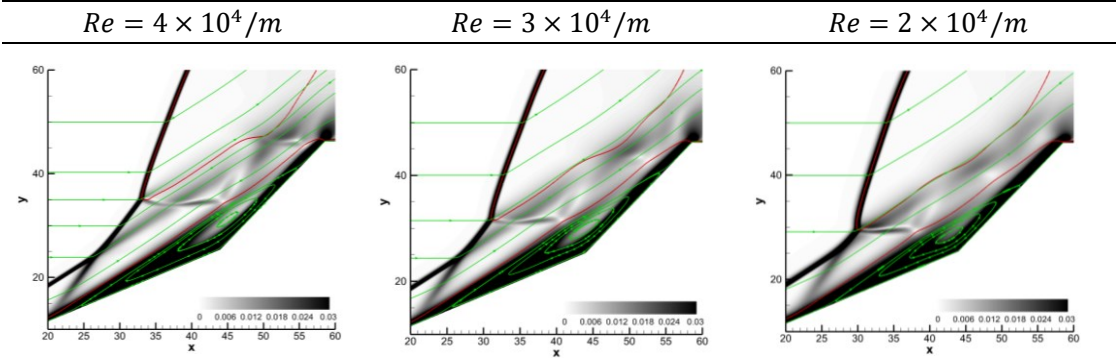

(a) Perfect gas

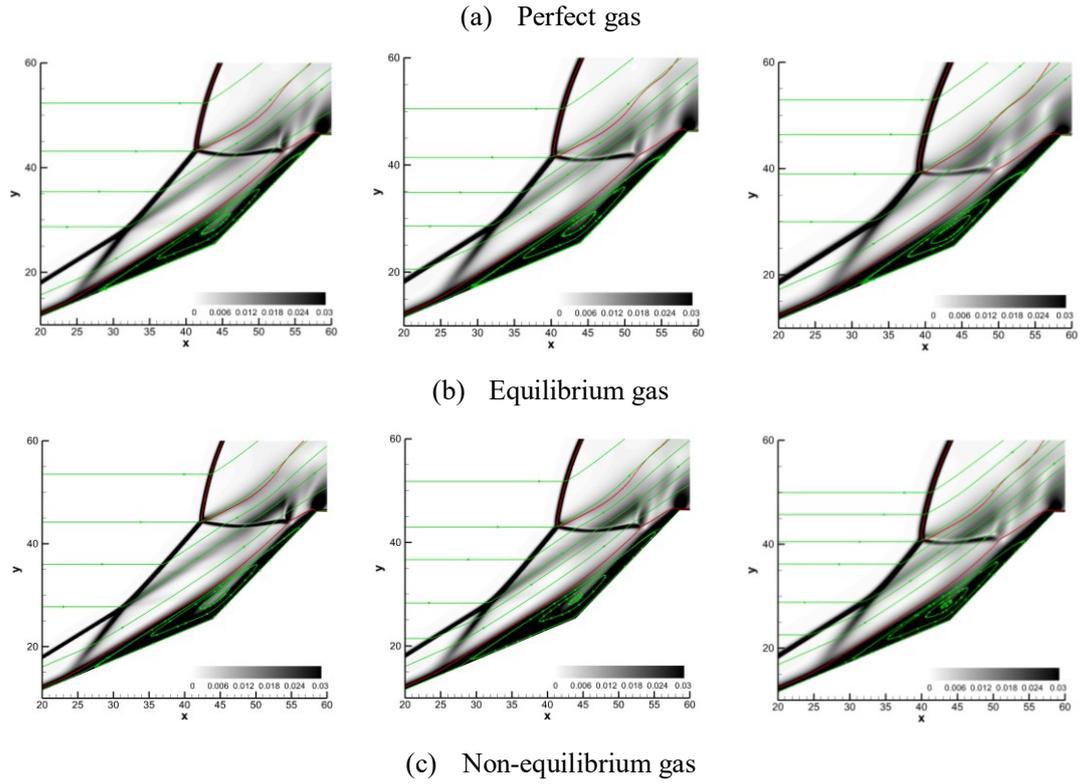

(b) Equilibrium gas

(c) Non-equilibrium gas

Fig. 19 Variations in vortex structure with $Re$ using three gas models at $Ma = 7$ and $h_0 = 2.1 MJ/kg$

To gain a deeper understanding of the nature of the interaction, a quantitative investigation is provided, and streamlines are chosen to run across the interaction structure, as shown in Fig. 18(a) by red dashed lines displaying variable distribution. However, due to the variation in interaction with $Re$, the lines do not converge into a single line. In Fig. 20, the pressure and $Ma$ distributions are shown along the chosen lines. Taking the case where $Re = 4 \times 10^4/m$ as an example, for illustration, the positions of LS, SS, TS, and RS, as well as the expansion wave (EW), are marked. The distributions reveal the following: (a) The values of pressure and $Ma$ remain nearly the same after LS, suggesting minimal impact from variations in $Re$. Similarly, SS exhibits weaker strength compared to LS; (b) The drop after EW verifies the reflection of TS (refer to Fig. 18(a)) as an expansion wave, which ends at the onset of compression waves around $x \approx 44.1mm$; (c) An abrupt increase in pressure at $x \approx 52.7mm$ signifies the occurrence of RS. Further analysis reveals $Ma$ of the flow component perpendicular to the wave to be 1.105, supporting the classification of CS as an oblique shock wave. A tertiary rise in pressure in the case $Re = 2 \times 10^4/m$ implies a further compression; (d) The distributions of $Ma$ are consistent to those of pressure, and further elaboration will be omitted for conciseness.

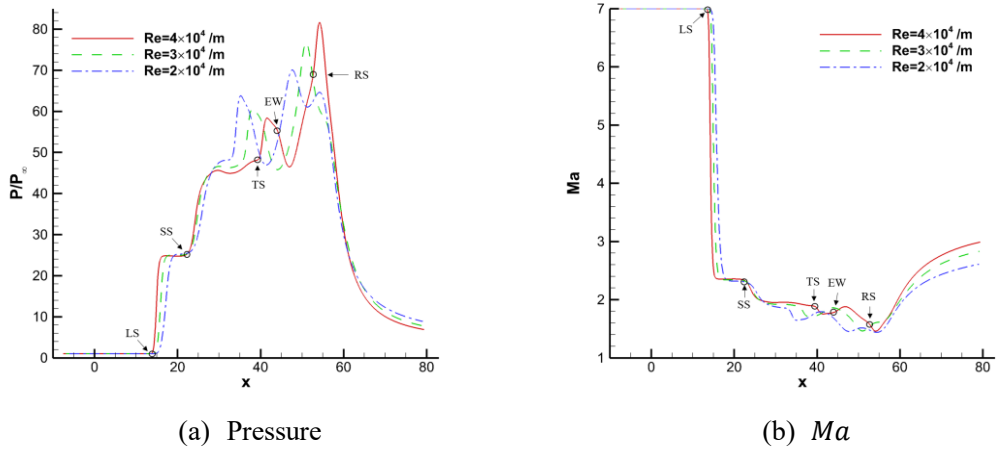

(a) Pressure  (b) $Ma$

Fig. 20 Variable distributions along the chosen lines across interaction structures of the perfect gas model such as transmitted shock, its reflection, etc. at three $Re$ under $Ma=7$ and $h_0=2.1 MJ/kg$

For comparison, similar distributions are shown in Fig. 21 for the non-equilibrium gas model, while the equilibrium gas model is not included due to its similarity to the former. The selected streamlines, depicted in Fig. 18(c) as red dashed lines, are chosen to have the same $y$=15 mm before LS at three $Re$. As a result, the lines are quite close to each other. For illustration, the positions of LS, SS, and TS, as well as EW, are marked in the distributions at $Re=4\times10^4/m$. The figures indicate that the distributions exhibit a high degree of similarity for the two larger $Re$, suggesting limited effects of $Re$, while some changes are observed at the smallest $Re$. By comparing the differences in distributions shown in Figs. 20, 21, and 14, it can be observed how the interactions between the perfect gas and real gas models perform and evolve differently.

In summary, the above study compares the variations in interactions of three gas models with $Re=4, 3,$ and $2\times10^4/m$. The interactions of the real gas models show similarities, with quantitative variations rather than fundamental changes. In contrast, the interactions of the perfect gas model exhibit a relatively drastic variation with $Re$ and differ from those of the real gas models.

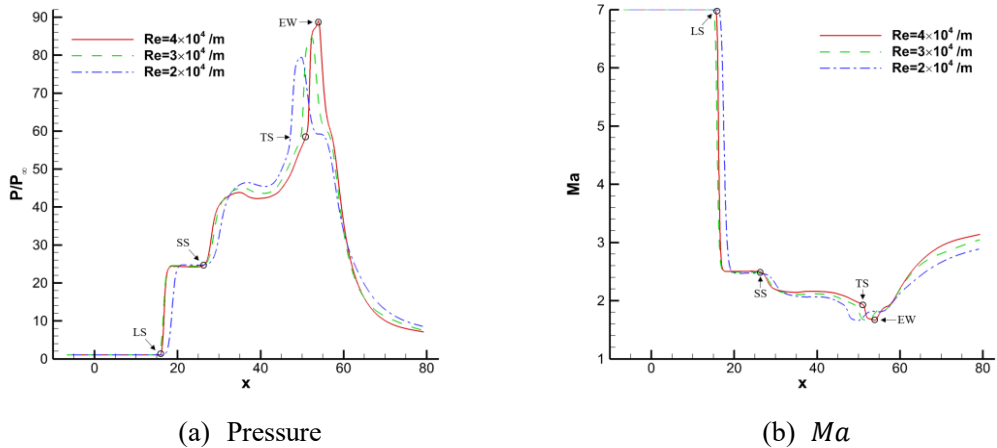

(a) Pressure  (b) $Ma$

Fig. 21 Variable distributions of the non-equilibrium gas model along the chosen lines with the same $y$=15 mm before LS and across interaction structures at three $Re$ under $Ma=7$ and $h_0=2.1 MJ/kg$

(1.3) Geometric characteristics and their variations with $Re$

Three types of characteristics are examined, i.e., those pertaining to primary shock waves, the triple point, and separation. To explore the asymptotic potential with decreasing $Re$, supplementary computations are performed at lower $Re$ till $Re = 1 \times 10^4/m$ with intervals of $0.5 \times 10^4/m$.

First, the shock angles of LS and SS with respect to the fore wedge are measured and shown in Fig. 22. The inviscid prediction of LS by the perfect gas model are provided for reference. Considering the dissipative nature of SS and the limitations in measurement accuracy, especially at small $Re$, some oscillations are observed. The figure indicates that for angles of LS, the viscous effect causes the numerical shock angles to be larger than the theoretical inviscid prediction, with the differences increasing further as $Re$ becomes less than $3.5 \times 10^4/m$. Among the available results of the three gas models, the perfect gas model shows a relatively larger prediction, especially for cases with small $Re$. This similar trend is also observed for the case of SS, which does not need to be reiterated.

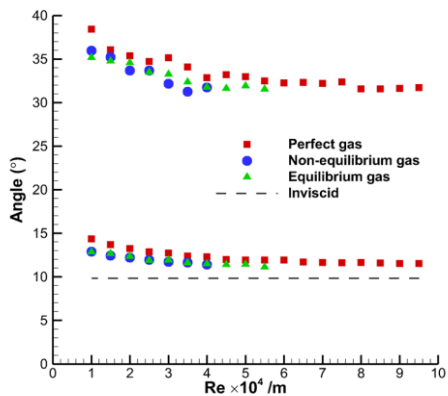 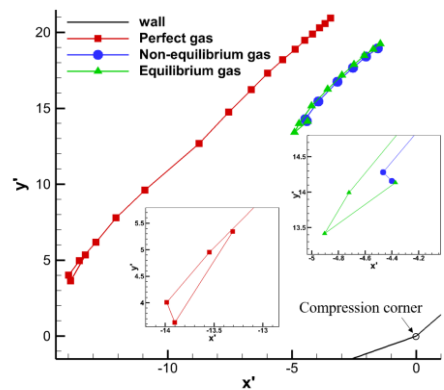

Fig. 22 Variation in shock angles of LS and SS with $Re$ of three gas models at $Ma = 7$ and $h_0 = 2.1 MJ/kg$ where inviscid prediction of LS by the perfect gas model is shown for reference

Fig. 23 Trajectories of triple point varying with $Re$ of three gas models at $Ma = 7$ and $h_0 = 2.1 MJ/kg$

Next, the geometric characteristics of triple points and their variations with $Re$ are examined. This includes analyzing the trace trajectories, the distances from the triple point to the compression corner, and the angles formed by the point, compression corner, and fore wedge. In order to better illustrate this, a coordinate transformation is performed by shifting the origin to the compression corner, as shown in Fig. 23. From the figure, it is evident that as $Re$ decreases, the triple points of the three models move upstream along the wedge. The triple point of the perfect gas model behaves differently from those of the real gas models, which exhibit similar trends. Moreover, in cases with the last two small $Re$, the movements appear to stall within the limits of current measurement accuracy, indicating that the interactions are approaching a consistent size rather than continuously shrinking with decreasing $Re$. Additionally, the other two geometric properties are shown in Fig. 24. Considering the measurement accuracy, it can be seen from the figure that similar to Fig. 23, the distributions of the distance and angle of the perfect gas model are clearly different from those of real gas models, which exhibit similar overall trends despite some differences. Meanwhile, variations in the properties also indicate rough convergence when $Re$ decreases to a small value; in addition, the distances of the three models converge toward similar values, while the angles of the perfect gas and real gas models converge differently.

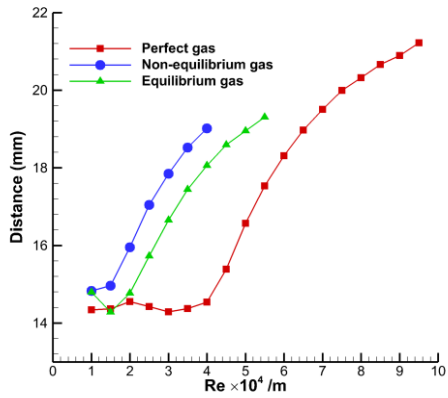 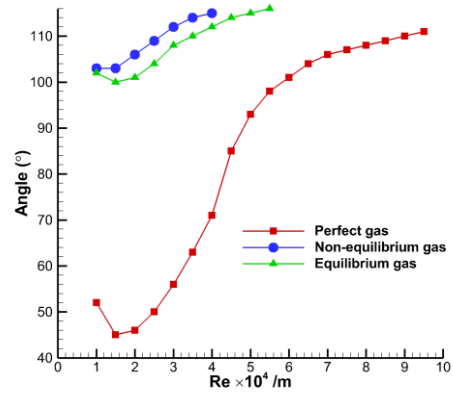

(a) Distance to compression corner  (b) Angle defined by triple point, compression corner and fore wedge

Fig. 24 Variation in geometric properties of triple point with $Re$ of three gas models at $Ma = 7$ and $h_0 = 2.1 MJ/kg$

Then, the geometric characteristics of separation, namely the distances of separation and reattachment points to the compression corner (referred to as $L_{SC}$, $L_{RC}$), the length of separation measured between these two points, and the angle of the separation streamline with respect to the fore wedge at the onset, are examined. More specifically, $L_{SC}$, the ratio of $L_{SC}/L_{RC}$, the length of separation, the angle, as well as their variations with $Re$ are measured, calculated, and shown in Fig. 25. In Fig. 25(a), it can be observed that all $L_{SC}$ exhibit a decreasing trend with decreasing $Re$, where $L_{SC}$ of the perfect gas model are notably larger than those of the real gas models, while the latter show similar values. Additionally, a plateau in the former model is evident at $Re = (3.5 - 4.5) \times 10^4/m$. Similar trends of variations in the separation length are observed as shown in Fig. 25(b), with the distributions of the three models being closer to each other, particularly for the perfect and equilibrium gas models at small $Re$. To further elucidate the characteristics, Fig. 25(c) illustrates the distribution of $L_{SC}/L_{RC}$, which indicates that the perfect gas model exhibits an overall increasing trend with decreasing $Re$ and has values greater than 1, whereas the real gas models show almost identical decreasing trends with values smaller than 1. The angles of separation are shown in Fig. 25(d), where a dashed line represents an angle of 12.5°. Assuming a separation having equal $L_{SC}$ and $L_{RC}$, the separation angle would be close to 12.5°. Therefore, Fig. 25(c) and (d) provide insights into the extent of deviation from symmetry in the predictions. It is evident that partial symmetries in terms of angle are somewhat achieved at $Re = (4 - 5) \times 10^4/m$. Moreover, a roughly increasing trend is observed with decreasing $Re$. By comparing Figs. 23–24 with Fig. 25, it is assumed that although the development of the triple point should be complete at the current low $Re$, the achievement of separation may not be feasible.

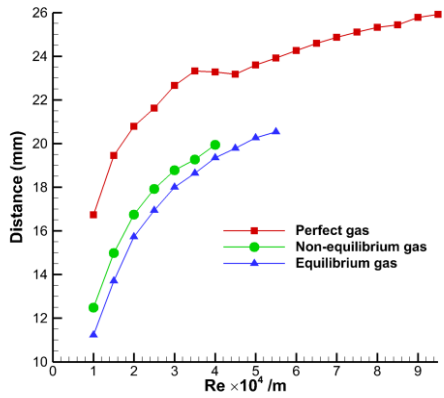
(a) $L_{SC}$

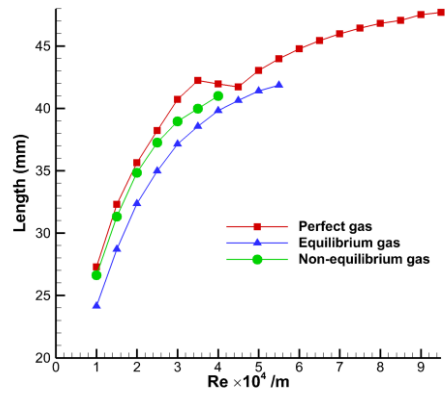
(b) Length of separation zone

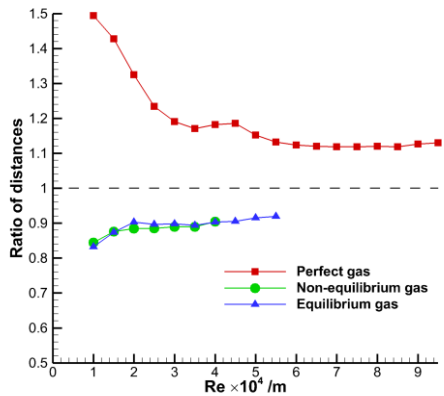
(c) $L_{SC}/L_{RC}$

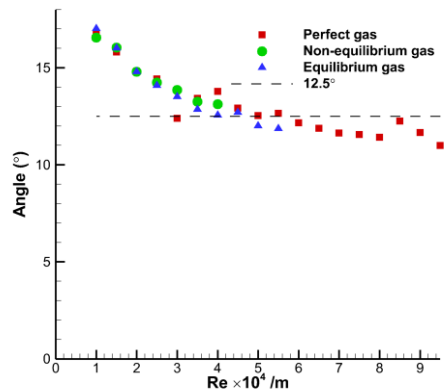
(d) Angle of separation streamline with respect to fore wedge

Fig. 25 Variation in geometric characteristics of separation zone with $Re$ of three gas models at $Ma = 7$ and $h_0 = 2.1 MJ/kg$

(2) Thermodynamic characteristics

As shown in the previous section, different gas models yield somewhat different performances, at least within a certain range of $Re$. Considering the case of low enthalpy and $Ma = 7$, it is plausible to suggest that the observed differences are primarily correlated to thermodynamic properties such as $\gamma$ and transport properties, which are usually closely linked to the temperature. Hence, we begin by examining the temperature distributions in different scenarios.

In Fig. 26(a), the temperature contours at the upper limit of $Re$ for the three models are displayed. The transmitted shocks impinge at locations closest to the compression corner. The figure shows that the perfect gas model yields a highest temperature of 2000 K, which theoretically could provoke real gas effects. In contrast, the high temperatures of the real gas models are notably lower at 1770–1800 K. Apart from the quantitative differences, the temperature distribution reveals four distinct regions. The first region is after the bow shock, where a concentration of high temperatures is observed. The second region is between the two slip lines, where the area after the transmitted shock has an even higher temperature. The third region is between the lower slip line and the separation. The fourth region is the separation zone and the area after the expansion corner, where lower temperatures are prominent, particularly near the wall, around the compression corner, and after the expansion corner.

The temperature contours vary with $Re = (4,3,2) \times 10^4/m$, as shown in Fig. 26(b) and (c) for the perfect gas and equilibrium gas models, respectively. The non-equilibrium case is not presented as it closely resembles the equilibrium situation. As $Re$ decreases, both sets of results exhibit a decrease or even disappearance of the second region. However, the perfect gas model shows an increased concentration in the first region, indicating intensified gas expansion and resulting in an enlarged bow shock. On the other hand, fewer variations are observed in the equilibrium gas case.

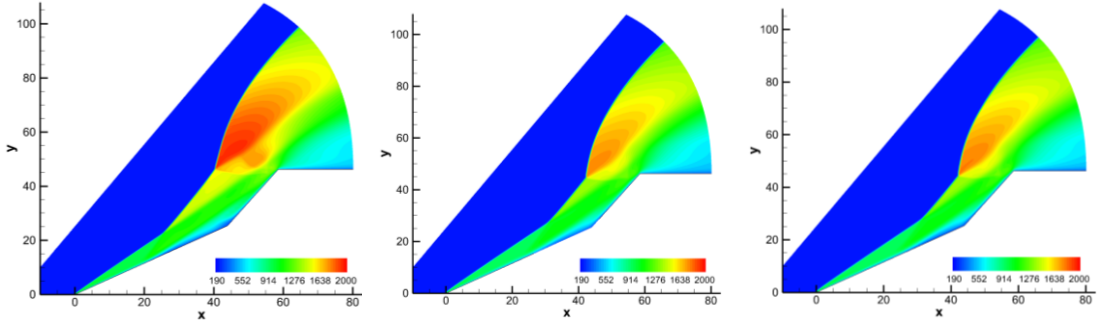

(a) Temperature contours at three upper $Re$ limits, namely $(9.5, 5.5, 4) \times 10^4/m$ of perfect, equilibrium and non-equilibrium gas models from left to right, respectively

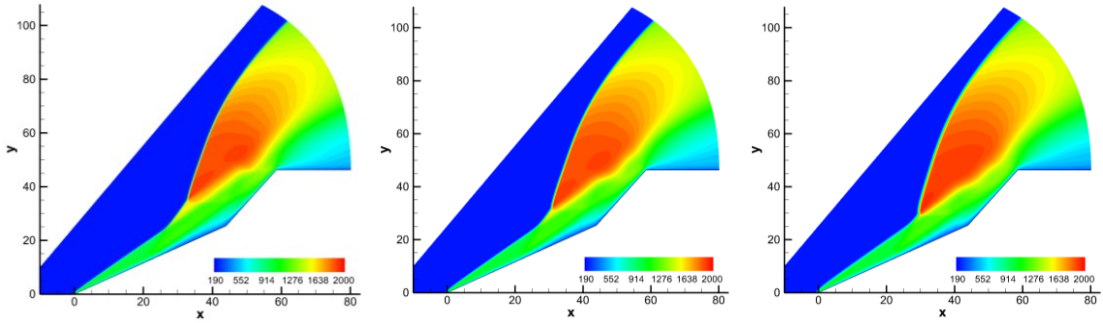

(b) Variations in temperature contours with $Re$ of perfect gas model (left to right: $Re = (4,3,2) \times 10^4/m$)

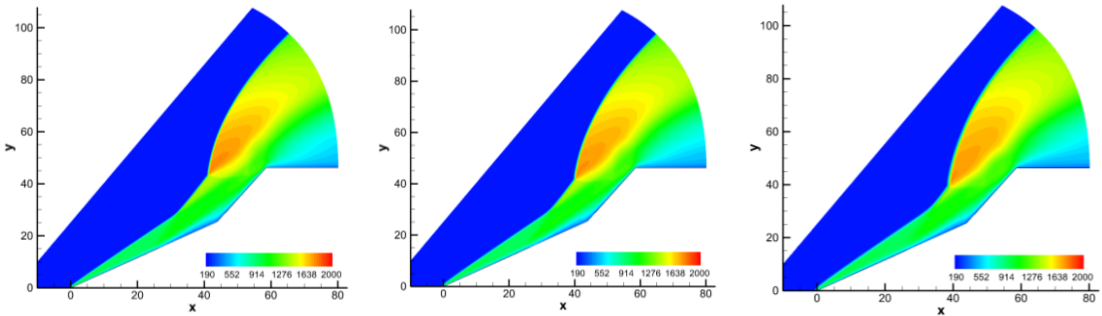

(c) Variations in temperature contours with $Re$ of equilibrium gas model (left to right: $Re = (4,3,2) \times 10^4/m$)

Fig. 26 Temperature contours and their variations with $Re$ at $Ma = 7$ and $h_0 = 2.1 MJ/kg$

Considering the temperature variation, it is natural to question the impact on thermodynamics. To this end, the specific heat ratio $\gamma$ is examined, and the condition $Re = 4 \times 10^4/m$ is chosen in view of the feasibility of steady solutions for real gas models. In Fig. 27(a) and (b), $\gamma$ contours for equilibrium ($\gamma_{EG}$) and non-equilibrium ($\gamma_{NEG}$) gas models are displayed, respectively, revealing values generally below 1.4. Specifically, in the first and second regions, $\gamma$ tends to have relatively smaller values, while in the fourth region near the wall, $\gamma$ begins to recover and approach 1.4.

Comparatively, $\gamma_{EG}$ is generally larger than $\gamma_{NEG}$. To explore these outcomes from another perspective, two additional quantities are derived, i.e. the percentage of $(\gamma_{EG} - \gamma_{NEG})/\gamma_{NEG}$ and $(1.4 - \gamma_{NEG})/1.4$, with corresponding contours shown in Fig. 27(c) and (d). Fig. 27(c) highlights that the differences between the two models are most prominent above the separation region, at around 3.5%. However, the differences become less noticeable around the separation and after the expansion corner. The tiny blue region ahead of BS in Fig. 27(c) arises due to the differences in BS shapes. Fig. 27(d) indicates that the relative difference of $\gamma_{NEG}$ with respect to 1.4 is over 6% after CS and even more after BS, with a level of approximately 5% between the two slip lines. The green and green-like areas represent a percentage of about 3—4%, while the region below 1.5% is confined to areas close to the wall, near the compression corner, and after the expansion corner.

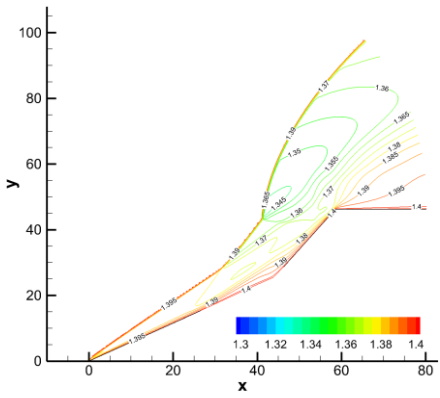

(a) $\gamma_{EG}$

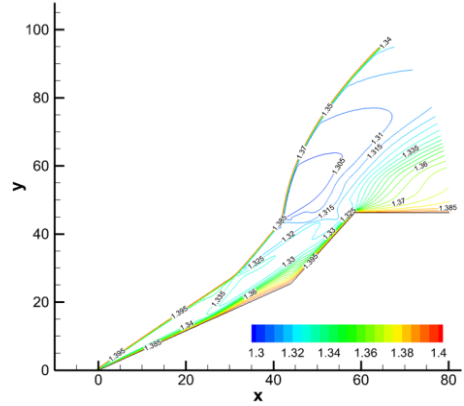

(b) $\gamma_{NEG}$

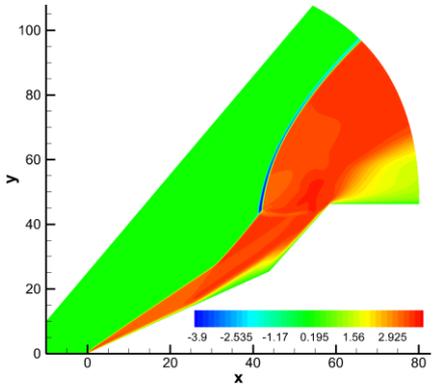

(c) $(\gamma_{EG} - \gamma_{NEG})/\gamma_{NEG} \times 100\%$

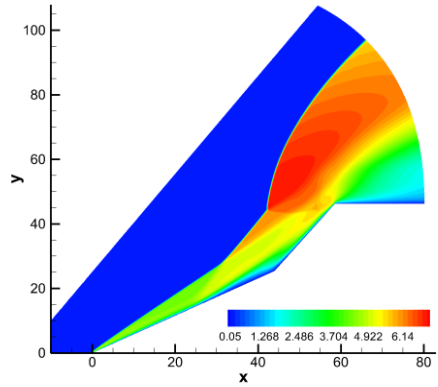

(d) $(1.4 - \gamma_{NEG})/1.4 \times 100\%$

Fig. 27 Contours of specific heat ratio of equilibrium and non-equilibrium gas models as well as relative differences at $Re = 4 \times 10^4/m$, $Ma = 7$, and $h_0 = 2.1 MJ/kg$

(3) Aerodynamic predictions along the wedge surface

Heat transfer and pressure distributions are the most common predictions made in engineering. As mentioned in the introduction, heat transfer is the main focus of scientific communities. It is well-known that heat transfer decreases with a reduction in $Re$ To draw a formal comparison with the experimental results of M7_2 [2, 3], it is common practice to present the data in different units in the same figure. This approach is also taken here, with the heat transfer values of the three models shown in Fig. 28. In Fig. 28(a)–(c), the $Re$ are at their upper limits besides $(4,3,2) \times 10^4/m$.

From the figure, the following observations can be made: (a) As expected, the order of heat transfer is much lower than that in the M7_2 experiment due to the small $Re$. As a result, the peak distributions at the interactions, e.g., TS impingement and reattachment locations, are too weak to be clearly visible. However, there is a noticeable abrupt attenuation of the boundary layer at the expansion corner, leading to a prominent heat peak. (b) For each model, the heat flux decreases as $Re$ decreases. As indicated in Figs. 18–19, SS is clearly located before the separation onset and dissipates at $Re = (4,3,2) \times 10^4/m$, resulting in a moderate decrease in heat transfer before separation onset. At higher $Re$ such as $9.5 \times 10^4/m$ in the perfect gas case, SS is more compact and therefore, a relatively steeper distribution is observed. (c) A comparison of the three models is made at $Re = 4 \times 10^4/m$, as shown in Fig. 28(d). The heat transfer of the perfect gas model shows an upstream shift of SS and separation onset, while in the real gas model, these points are closer together. Additionally, the heat transfer in the former model is higher than that in the latter at around 0.054 mm, suggesting complex interactions. The close/same distributions before SS indicate that the effect of the gas model is insignificant in that region.

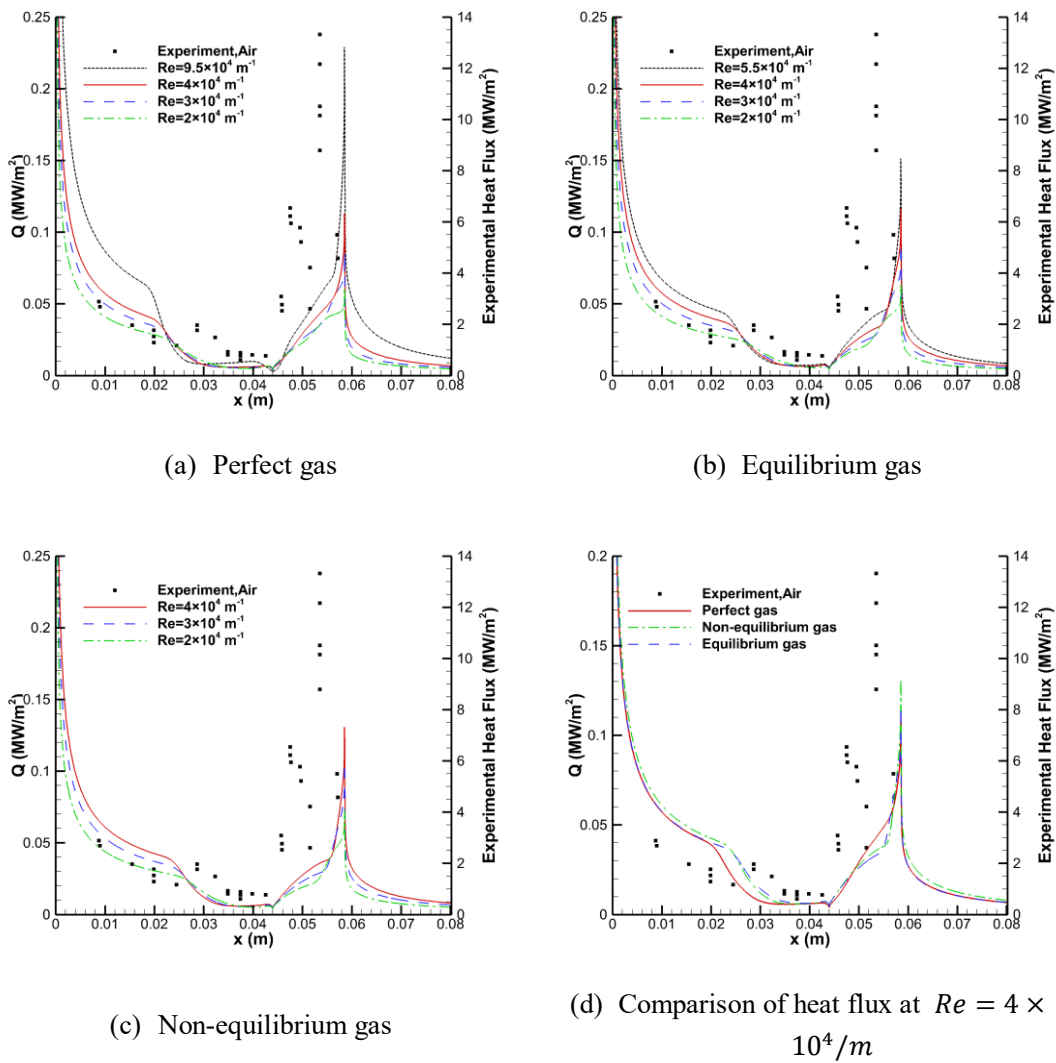

(a) Perfect gas

(b) Equilibrium gas

(c) Non-equilibrium gas

(d) Comparison of heat flux at $Re = 4 \times 10^4/m$

Fig. 28 Heat flux distributions under several $Re$ of three gas models with the reference of M7_2 [2, 3] and comparison of three models at $Re = 4 \times 10^4/m$, $Ma = 7$, and $h_0 = 2.1 MJ/kg$

Similarly, the pressure distributions are examined and displayed in Fig. 29. For reference, the inviscid predictions by the perfect and equilibrium gas models are indicated using a double dot-dash line. It can be seen from Fig. 29(a)–(c) that all models achieve consistent predictions of pressure along the fore wedge before SS at the chosen $Re$, indicating that the effect of $Re$ is negligible in that region. Furthermore, good agreement is obtained with the inviscid prediction, except near the wedge apex, which demonstrates satisfactory numerical accuracy. However, disparities are observed in the region after SS, especially at the separation zone, where a smaller $Re$ leads to a larger pressure coefficient, implying the effects of viscosities in that area. In Fig. 29(d), the results of different models at $Re = 4 \times 10^4/m$ are compared, and the figure shows little differences before SS, indicating an insignificant effect of the real gas models. However, the perfect gas model shows an early occurrence of SS and a larger pressure at the separation zone, while the results of the real gas models closely resemble each other.

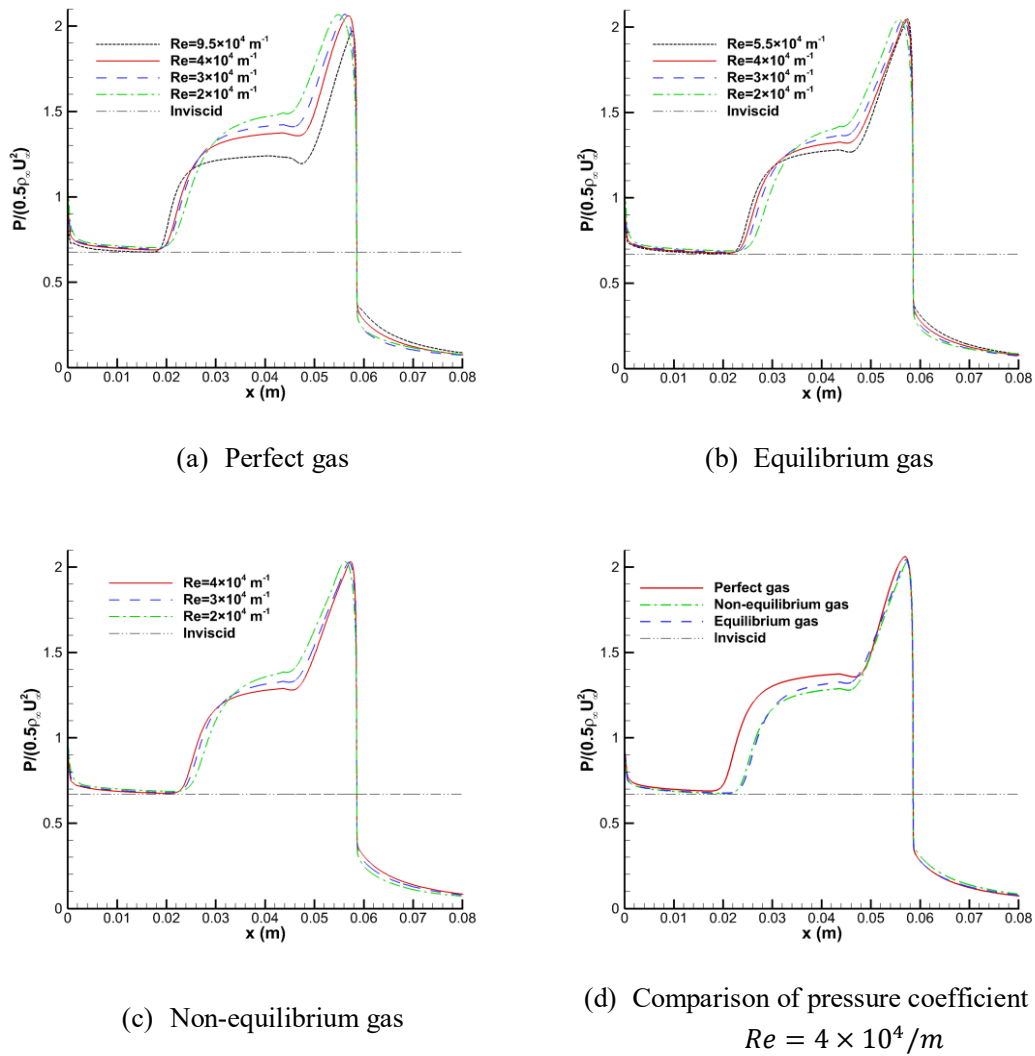

(a) Perfect gas

(b) Equilibrium gas

(c) Non-equilibrium gas

(d) Comparison of pressure coefficient at $Re = 4 \times 10^4/m$

Fig. 29 Distributions of pressure coefficient at chosen $Re$ of three gas models and comparison of three models at $Re = 4 \times 10^4/m$ under $Ma = 7$ and $h_0 = 2.1 MJ/kg$ with the reference of inviscid prediction

## 5 Analysis of interaction pattern and quasi-normal shock in the case of the perfect gas model

In the previous section, the steady interactions of a double wedge were studied in detail at various $Re$ under $Ma = 7$ and $h_0 = 2.1 MJ/kg$. The maximum $Re$ at which the flows of three models can remain steady are found to be $(9.5, 5.5, 4) \times 10^4/m$. As indicated in [26], the shock polar can serve as a useful measure to understand interaction mechanisms, and similar practices are employed in this section. Prior to the main part of the analysis, some preliminary discussions will be provided: (a) The wall condition plays a crucial role in the shock polar for determining the flow direction, with the inviscid wall usually chosen for simplicity [27]. In the current study, the boundary layers cannot be ignored due to the modest $Re$; hence, predictions based on the inviscid wall may differ from the viscous results. Moreover, when TS impinges on the separation zone, the corresponding separation line is treated as a virtual wall, which may lead to inaccuracies. (b) When comparing the results of the non-equilibrium gas model, it appears reasonable to consider its effect in the shock polar analysis. However, the feasibility of this practice depends on the availability of relaxation time, which can be challenging to obtain in advance. To illustrate the potential differences in predictions between the two real gas models, the pressure and temperature distributions are compared in Fig. 30 along the chosen lines at $y$=15 mm before LS, as shown by the dashed lines in Fig. 18(b) and (c) at $Re = 4 \times 10^4/m$ under $Ma = 7$ and $h_0 = 2.1 MJ/kg$. From the figure, the distributions of the two models are almost identical, suggesting that the relaxation process may be trivial. Considering the relatively low $Ma$ in this study, it is reasonable to use the shock polar based on the equilibrium model to analyze the results of the non-equilibrium gas model. In summary, two types of shock polar methods are employed: the canonical one for the perfect gas model to analyze results within the same model, and the one based on the equilibrium gas model for results of the two real gas models. Next, the chosen cases are investigated.

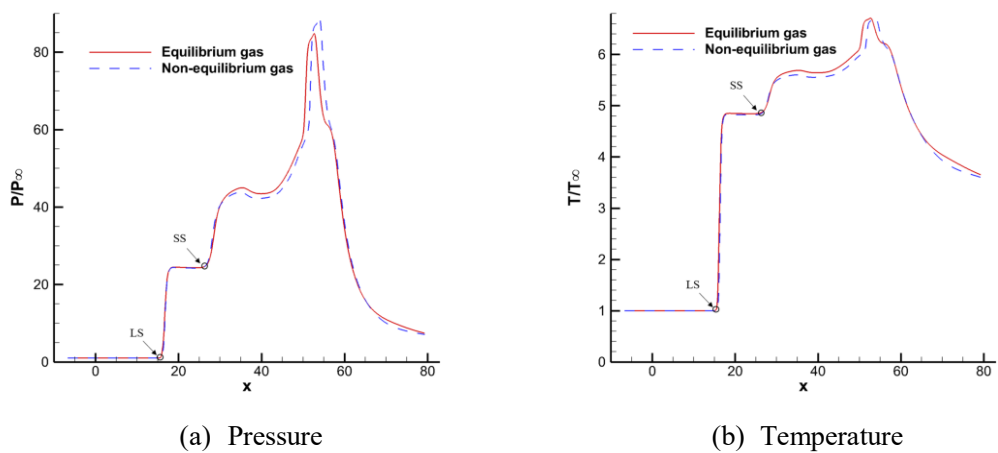

(a)  Pressure  (b)  Temperature

Fig. 30 Comparisons of variable distributions of real gas models along the chosen lines with the same $y$=15 mm before LS and across interaction structures at $Re = 4 \times 10^4/m$ under $Ma = 7$ and $h_0 = 2.1 MJ/kg$

(1) Perfect gas case at $Re = 9.5 \times 10^4/m$ under $Ma = 7$ and $h_0 = 2.1 MJ/kg$

As previously mentioned, in this case, $Re$ reaches the upper limit of the perfect gas model to achieve a steady flow. Simplifying the interactions and neglecting minor effects, a sketch can be created based on the numerical results discussed in Section 4. Typical regions are numbered as shown in Fig. 31(a), where "SL" denotes the slip line. Meanwhile, corresponding shock polar curves are presented in Fig. 31(b) by plotting $p/p_\infty$ against the deflection angle, where the anticlockwise direction is considered negative. The states at each region are determined as follows: The state at Region 1 is defined by the shock polar in black line under freestream conditions at a fore wedge angle of 30°. The state at Region 2 is defined by the shock polar of region 1 (scarlet line) at the angle of separation, namely 40°, as determined by the numerical simulation. Region 3's state is established by Prandtl–Meyer expansion at the location where the expansion in the dashed line intersects with the freestream polar curve in black. Region 4's state is defined as the one on shock polar of the freestream at the intersection. States at Regions 5 and 6 are derived from the intersection of the shock polar from region 4 (purplish red line) with the freestream polar in black, taking into account the upper slip line. States at Regions 7 and 8 are determined by the interaction of the shock polar curve of Region 3 (blue line) with that at Region 4, considering the lower slip line. It is important to note that the angle of separation, 40°, is defined by averaging the angles of the separation line near the separation point due to the fact that the line changes slightly in that area, leading to corresponding variations in the angle.

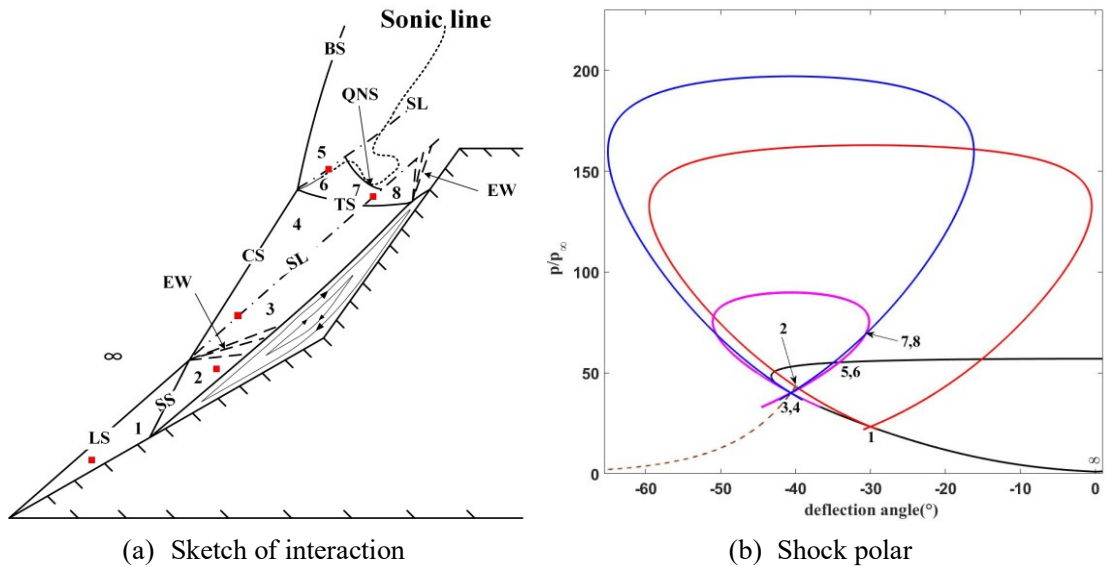

(a) Sketch of interaction        (b) Shock polar

Fig. 31 Sketch of interaction and corresponding shock polar of the case of the perfect gas model at $Re = 9.5 \times 10^4/m$ under $Ma = 7$ and $h_0 = 2.1 MJ/kg$

Based on the analysis above, the following comparisons are made. First, five points shown in red squares in Fig. 31(a) are chosen as representatives, with their coordinates and associated regions listed in Table 2. The measured pressure ratio is then compared with the predictions from shock polar analysis, as shown in the table. The relative difference between the measurement and prediction with respect to the former is also computed. The table indicates that in the current perfect gas case, the shock polar method shows reasonable agreement with the computations at the chosen points, suggesting the validity and capability of the method in the presence of viscosity. It is also important to note that the angle of separation is determined through computation, indicating that the prediction is not entirely closed-loop.

Table 2. Coordinates and affiliated regions of chosen points and pressure ratios obtained by predictions and measurements for $Re = 9.5 \times 10^4/m$ with $Ma = 7$ and $h_0 = 2.1 MJ/kg$

| Point | Coordinates | Associated region | $r_p = p/p_\infty$ Shock polar (sp) | $r_p = p/p_\infty$ Measured (m) | $\frac{|r_{p,m} - r_{p,sp}|}{r_{p,m}}$ (%) |
|---|---|---|---|---|---|
| P1 | (15.67, 12.07) | 1 | 23.33 | 24.30 | 3.99% |
| P2 | (28.53, 21.06) | 2 | 43.26 | 42.48 | 1.84% |
| P3 | (32.56, 28.50) | 3&4 (slip line) | 40.06 | 40.70 | 1.57% |

| P4 | (43.10, 48.18) | 5&6 (slip line) | 55.20 | 54.77 | 0.79% |
| P5 | (51.54, 45.49) | 7&8 (slip line) | 68.90 | 65.89 | 4.57% |

(2) Equilibrium gas case at $Re = 5.5 \times 10^4/m$ under $Ma = 7$ and $h_0 = 2.1 MJ/kg$

Likewise, $Re$ in this case reaches an upper limit at which the equilibrium gas model can achieve a steady flow. A sketch of the interaction is shown in Fig. 32(a), where the angle of separation defining the state at Region 2 is measured as 40°. The analysis for a similar configuration is completely the same as that described above, and the corresponding shock polar curves are shown in Fig. 32(b) using the same indices to indicate intersections. For brevity, repetition is avoided.

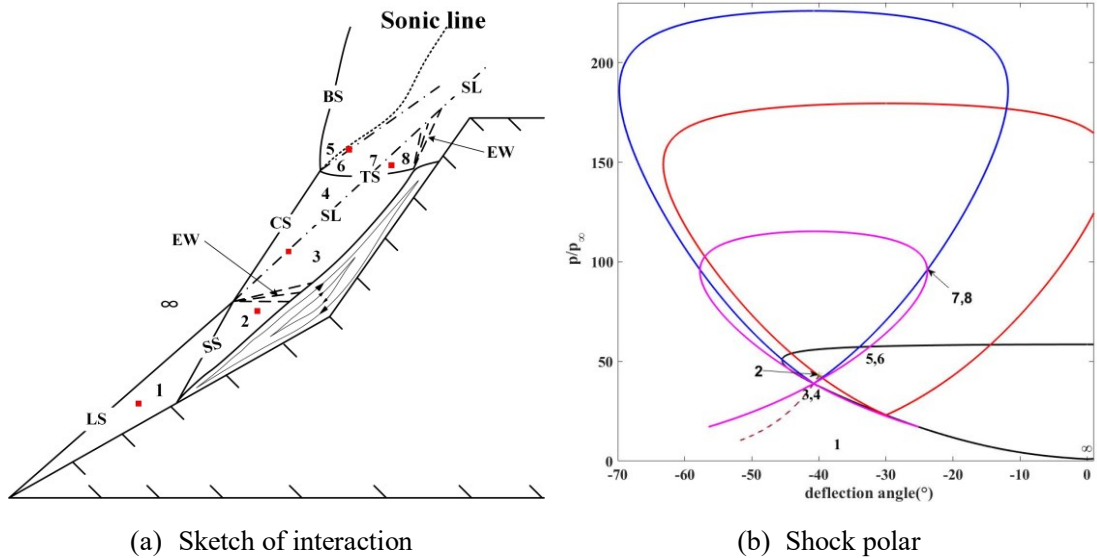

(a) Sketch of interaction    (b) Shock polar

Fig. 32 Sketch of interaction and corresponding shock polar of the case of the perfect gas model at $Re = 5.5 \times 10^4/m$ under $Ma = 7$ and $h_0 = 2.1 MJ/kg$

For comparison, five similar representative points are chosen, as shown in Fig. 32(a), and the results are presented in Table 3, similar to Table 2. It is worth mentioning that the thermodynamic relations in the predictions are based on the equilibrium gas model introduced in Section 2.1. The outcomes differ from those obtained using the perfect gas model to a certain extent. The results indicate slightly larger differences between the measurements and predictions, possibly owing to the indirect influence of the gas model on the viscous formations such as boundary layers and separations.

Table 3. Coordinates and affiliated regions of chosen points and pressure ratios obtained by predictions and measurements for $Re = 5.5 \times 10^4/m$ with $Ma = 7$ and $h_0 = 2.1 MJ/kg$

| Point | Coordinates | Associated region | $r_p = p/p_\infty$ | | $\dfrac{\lvert r_{p,m} - r_{p,sp} \rvert}{r_{p,m}}$ (%) |
| --- | --- | --- | --- | --- | --- |
| | | | Shock polar (sp) | Measured (m) | |
| P1 | (15.67, 12.07) | 1 | 23.06 | 24.30 | 5.10% |
| P2 | (32.00, 24.23) | 2 | 42.88 | 42.96 | 0.19% |
| P3 | (36.44, 32.02) | 3&4 (slip line) | 38.91 | 40.91 | 4.89% |
| P4 | (45.08, 46.44) | 5&6 (slip line) | 57.46 | 58.68 | 2.08% |
| P5 | (53.86, 45.53) | 7&8 (slip line) | 95.94 | 89.46 | 7.49% |

(3) Cases of three models at $Re = 4 \times 10^4/m$ under $Ma = 7$ and $h_0 = 2.1 MJ/kg$

As mentioned earlier, the non-equilibrium gas model can achieve a steady flow at the highest $Re$, while the perfect gas model shows a unique interaction pattern. In the following section, the cases of both the perfect gas model and real gas models are discussed.

(3.1) Perfect gas model

A sketch of the interaction is shown in Fig. 33(a). As previously discussed, a reflection of TS occurs at the separation, following EW, CW, and RS. Due to the increased complexity, the uncertainty of predictions intensifies accordingly. Therefore, the analysis using the shock polar method still stops behind BS and TS, as it did previously. Upon comparing Fig. 33(a) with Figs. 32(a) and 31(a), a simplification can be observed: the expansion wave is absent in the interaction between LS and SS, leading to a more straightforward prediction. This simplification is implemented because in Fig. 19(a), the first column indicates an indiscernible CW, while Fig. 20(a) shows a trivial pressure expansion. When observing the numbered regions in Fig. 33(a), the process can be described as follows: The state at Region 1 is defined by the shock polar in a black line of the freestream condition at the angle of the fore wedge. The state at Region 2 is defined by the shock polar of region 1 in a scarlet line at an angle of separation measured as 42°, while Region 3's state is determined on the shock polar of the freestream at the same angle as Region 2. Subsequently, the states at Regions 4 and 5 are derived by intersecting the shock polar of Region 3 in a purplish red line with that of the freestream in a black line, and the state at Region 6 is defined by the intersection of the shock polar of Region 3 in a purplish red line with the shock polar of Region 2 in a blue line.

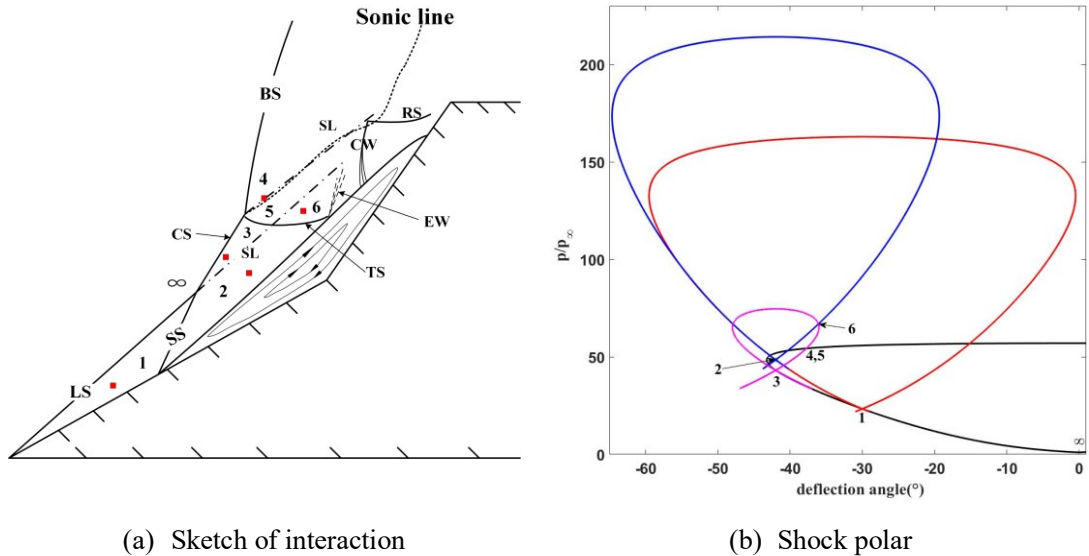

(a) Sketch of interaction  (b) Shock polar

Fig. 33 Sketch of interaction and corresponding shock polar of the case of perfect gas model at $Re =\times 10^4/m$ under $Ma = 7$ and $h_0 = 2.1 MJ/kg$

Similarly, five representative points are chosen as indicated in Fig. 33(a). The coordinates and corresponding pressures measured and predicted by the perfect gas model are presented in Table 4. A comparison between the measurements and predictions reveals satisfactory agreements, with the exception of a relatively larger difference observed at P5 or at Region 6.

Table 4. Coordinates and affiliated regions of chosen points and their pressure ratios predicted by the perfect gas model and measurements for $Re = 4 \times 10^4/m$ with $Ma = 7$ and $h_0 = 2.1 MJ/kg$

| Point | Coordinates | Associated region | $r_p = p/p_\infty$ | | $\frac{|r_{p,m} - r_{p,sp}|}{r_{p,m}}$ (%) |
|---|---|---|---|---|---|
| | | | Shock polar (sp) | Measured (m) | |
| P1 | (15.67, 12.07) | 1 | 23.33 | 24.78 | 5.85% |
| P2 | (30.90, 24.94) | 2 | 48.50 | 45.64 | 6.27% |
| P3 | (32.94, 32.21) | 3 | 43.16 | 44.47 | 2.94% |
| P4 | (34.92, 36.63) | 4&5 (slip line) | 54.29 | 55.18 | 1.61% |
| P5 | (39.64, 35.06) | 6 | 66.84 | 59.03 | 13.23% |

(3.2) Real gas model

In the case of real gas models, the interactions are quite similar to each other as discussed in Section 4. Upon examination, it was found that their sketches and shock polar curves are qualitatively identical to Fig. 32 in "(2)". Therefore, the mechanisms follow suit, and no further discussions will be provided, except for noting that the angle of separation is found to be 41° for both models. To facilitate comparison, five similar points are chosen as those shown in Fig. 32(a), albeit with different coordinates. Consequently, the pressure ratios at these points, as predicted by the equilibrium gas model and measurements are presented and compared in Table 5. In the table, the subscript "EQ" indicates points corresponding to the equilibrium gas model, while "NEQ" denotes those for the non-equilibrium model. Despite some discrepancies, the comparisons reveal a reasonable level of agreement.

Table 5. Coordinates and affiliated regions of chosen points and their pressure ratios predicted by the real gas models and measurements for $Re = 4 \times 10^4/m$ with $Ma = 7$ and $h_0 = 2.1 MJ/kg$

| Point | Coordinates | Associated region | $r_p = p/p_\infty$ | | $\frac{|r_{p,m} - r_{p,sp}|}{r_{p,m}}$ (%) |
|---|---|---|---|---|---|
| | | | Shock polar (sp) | Measured (m) | |
| P1$_{EQ}$ | (15.67, 12.07) | 1 | 23.08 | 24.50 | 5.80% |
| P2$_{EQ}$ | (30.28, 21.95) | 2 | 45.41 | 42.62 | 6.55% |
| P3$_{EQ}$ | (35.95, 32.05) | 3&4 (slip line) | 40.80 | 41.73 | 2.23% |
| P4$_{EQ}$ | (43.49, 44.69) | 5&6 (slip line) | 57.18 | 58.75 | 2.67% |
| P5$_{EQ}$ | (52.47, 44.18) | 7&8 (slip line) | 87.48 | 86.07 | 1.64% |
| Point | Coordinates | Affiliated region | $r_p = p/p_\infty$ | | $\frac{|r_{p,m} - r_{p,sp}|}{r_{p,m}}$ (%) |
| | | | Shock polar (sp) | Measured (m) | |
| P1$_{NEQ}$ | (15.67, 12.07) | 1 | 23.08 | 24.41 | 5.45% |
| P2$_{NEQ}$ | (30.28, 21.95) | 2 | 45.41 | 42.12 | 7.81% |
| P3$_{NEQ}$ | (37.36, 33.12) | 3&4 (slip line) | 40.80 | 41.21 | 2.74% |
| P4$_{NEQ}$ | (43.95, 45.21) | 5&6 (slip line) | 57.18 | 59.36 | 3.61% |
| P5$_{NEQ}$ | (53.56, 45.21) | 7&8 (slip line) | 87.48 | 89.37 | 2.11% |

(4) Mechanism of quasi-normal shock of perfect gas model

In Section 4, a quasi-normal shock is reported for the case where $Re = 9.5 \times 10^4/m$, $Ma = 7$ and $h_0 = 2.1 MJ/kg$, within a passage confined by two slip lines. The variable distributions along the streamline across the middle of QNS are analyzed. A mechanism is proposed, as illustrated in Fig. 34, to explain the generation of QNS. Specifically, the presence of two converging slip lines causes the supersonic flow within the passage to undergo compression and isentropic deceleration, until the flow reaches a point where it can no longer maintain its supersonic speed. At this point, a shock is engendered to match the flux and speed. Subsequently, the flow undergoes an isentropic process, which is primarily controlled by the passage area.

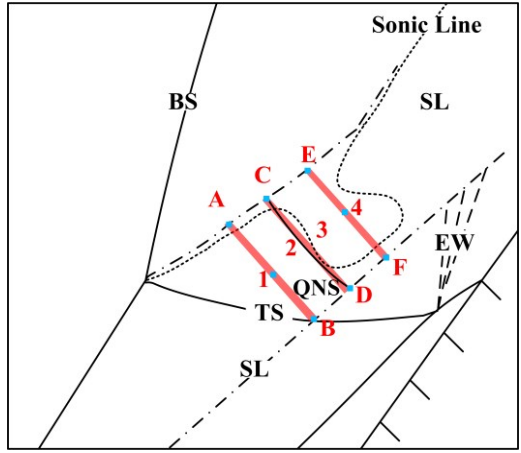

Fig. 34 Sketch of quasi-normal shock of the perfect gas model at $Re = 9.5 \times 10^4/m$ under $Ma = 7$ and $h_0 = 2.1 MJ/kg$

To analyze such a process, a 1D model is used in which the difference normal to the streamwise direction is ignored. Given the area ($A$) and $Ma$ at position 1, as well as the area and pressure at position 4, supposing 2 and 3 to be the front and back of the shock with $A_{23}$ to be determined, the ratio of $A$ and pressure at positions 1-2 and 3-4 satisfy the following relations in terms of $Ma$ under the isentropic process:

$$\begin{cases} \frac{A_{23}}{A_1} = \frac{Ma_1}{Ma_2} \times \left(\frac{1+(\gamma-1)Ma_2^2/2}{(\gamma-1)Ma_1^2/2}\right)^{\frac{\gamma+1}{2(\gamma-1)}}, \frac{A_4}{A_{23}} = \frac{Ma_3}{Ma_4} \times \left(\frac{1+(\gamma-1)Ma_4^2/2}{(\gamma-1)Ma_3^2/2}\right)^{\frac{\gamma+1}{2(\gamma-1)}} \\ \frac{p_1}{p_2} = \frac{Ma_1}{Ma_2} \times \left(\frac{1+(\gamma-1)Ma_2^2/2}{(\gamma-1)Ma_1^2/2}\right)^{\frac{\gamma}{(\gamma-1)}}, \frac{p_3}{p_4} = \frac{Ma_3}{Ma_4} \times \left(\frac{1+(\gamma-1)Ma_4^2/2}{(\gamma-1)Ma_3^2/2}\right)^{\frac{\gamma}{(\gamma-1)}} \end{cases}.$$

Meanwhile, the shock relations give

$$\begin{cases} \frac{p_3}{p_2} = \frac{1+\gamma \times Ma_2^2}{1+\gamma \times Ma_3^2} \\ Ma_3^2 = \frac{2/(\gamma-1)+Ma_2^2}{2\gamma/(\gamma-1) \times Ma_2^2 - 1} \end{cases}.$$

Thus, $A_{23}$ can be solved by the above equations.

In the current situation, the area of the passage can be expressed by the width, as indicated in Fig. 34. The required inputs are obtained from the computation as follows: $A_1 = 5.163mm$, $A_4 = 5.524mm$. The pressure and $Ma$ on chosen points, as shown in Fig. 34, are $p_1/(\rho_\infty u_\infty^2) = 0.781$, $p_4/(\rho_\infty u_\infty^2) = 1.015$, and $Ma_1 = 1.1950$. Based on the inputs and the above equations, $A_{23}$ can be solved as $A_{23} = 5.01mm$, while the measured value is $5.14mm$. Therefore, the result of computation agrees well with that predicted by the 1D model, and the mechanism provided can be considered reasonable.

## 6 Effect of variations in Mach number and enthalpy

In Section 4, the steady interactions were comprehensively studied with $Re$ spanning over $(1 \sim 9.5) \times 10^4/m$ at $Ma = 7$ and $h_0 = 2.1 MJ/kg$. It is natural to wonder what would happen if $Ma$ or $h_0$ varies. Given the upper limits of $Re$ obtained, $Re$ is chosen as $4 \times 10^4/m$ here.

### 6.1 Effect of $Ma$ variation at $h_0 = 2.1 MJ/kg$ and $Re = 4 \times 10^4/m$

In this scenario, in addition to the original $Ma = 7$, two more $Ma$ are considered, namely 8 and 9. For completeness, the inflow conditions are given as follows: $T_\infty = 191K, p_\infty = 14.44pa$ at $Ma = 7$; $T_\infty = 153.774K, p_\infty = 9.344pa$ at $Ma = 8$; and $T_\infty = 123.376K, p_\infty = 6.042$ at $Ma = 9$. It is worth noting that $T_\infty$ decreases as $Ma$ increases.

Computations are carried out using three models and steady results are achieved. For brevity, this paper mainly discusses the results obtained from the non-equilibrium gas model for representation. Moreover, the results at $Ma = 7$ are referenced in Section 4 and will not be repeated. In Fig. 35(a), the pressure contours of the non-equilibrium gas model are presented, with corresponding vortex structures displayed in Fig. 35(b). These figures indicate that there are no fundamental changes in interaction patterns, as detailed in Sections 4 and 5. However, closer examination reveals quantitative differences that will be discussed next. In addition to pressure distributions, temperature distributions are shown in Fig. 35(c) and in Fig. 26(a), reflecting a similar scenario as that of the pressure. It is worth noting that the absence of a significant temperature increase is attributed to the decrease in inflow temperature under constant $h_0$.

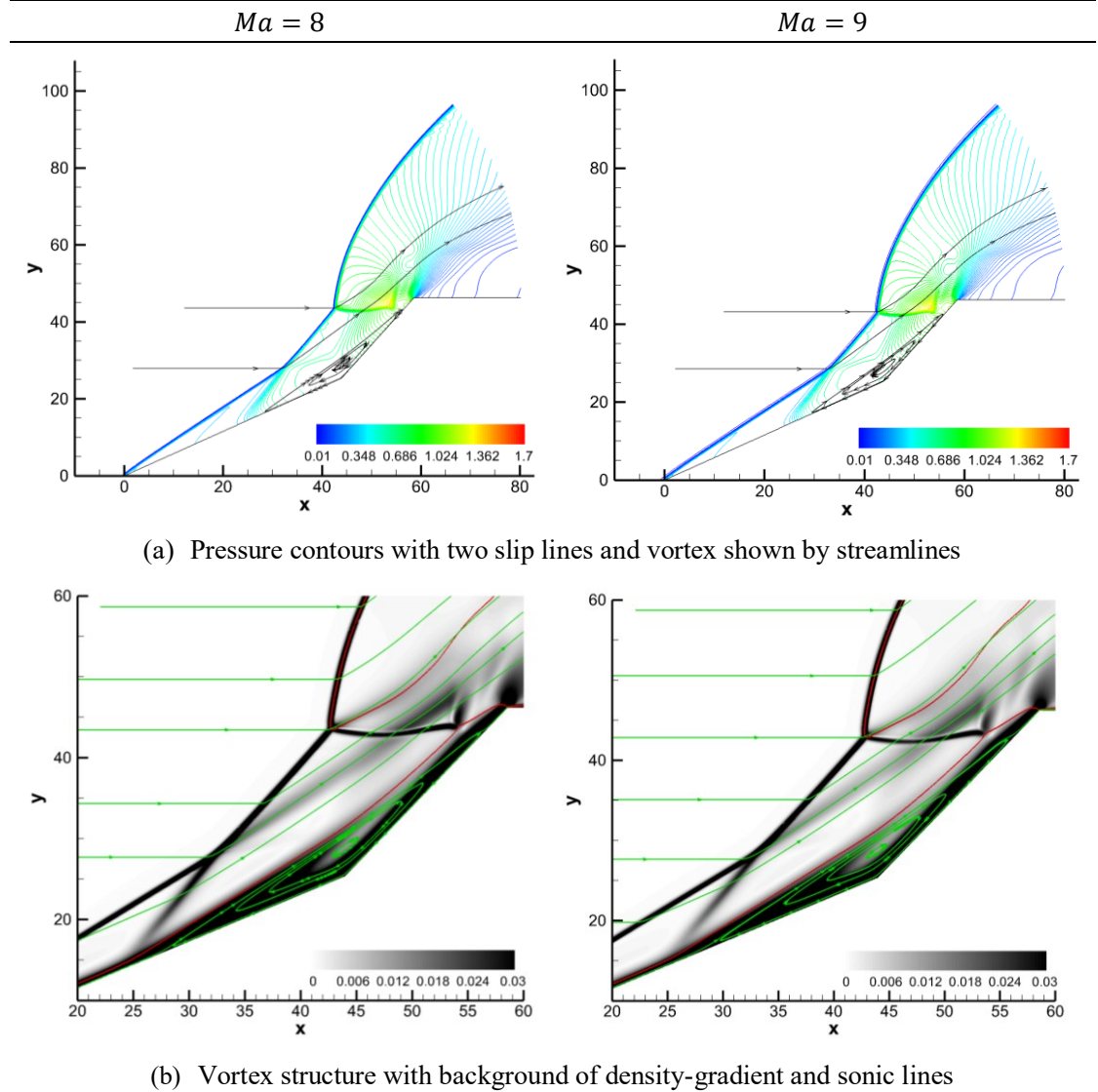

(a) Pressure contours with two slip lines and vortex shown by streamlines

(b) Vortex structure with background of density-gradient and sonic lines

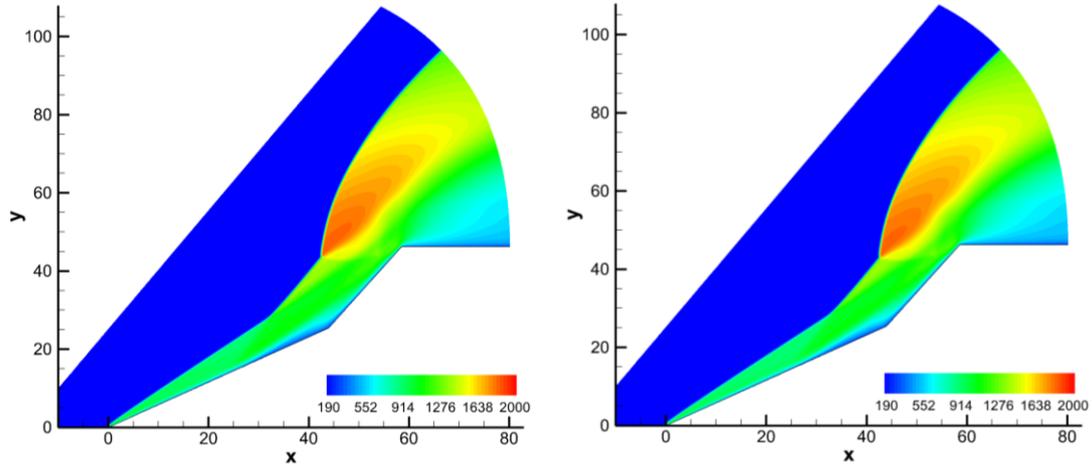

(c) Temperature contours

Fig. 35 Variations in variable contour and vortex structure of the non-equilibrium gas model with $Ma$ at $h_0 = 2.1 MJ/kg$ and $Re = 4 \times 10^4/m$

To quantitatively demonstrate the geometric characteristics, the trajectories of impingement and triple points in the non-equilibrium gas model are first analyzed, as shown in Fig. 36. The coordinate transformations mentioned earlier have been applied, and the arrows in the figure indicate the direction of $Ma$ increase. The impingement points appear to be moving away from either the expansion corner or the aft wedge with an increase in $Ma$, while an approximate shift of the triple points towards the compression corner is also observed.

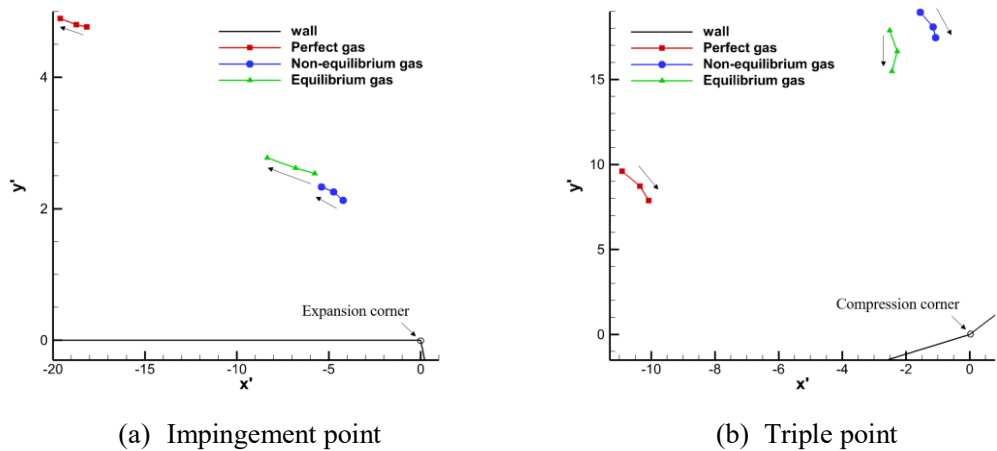

(a) Impingement point  (b) Triple point

Fig. 36 Trajectories of characteristic points of three gas models with $Ma$ at $h_0 = 2.1 MJ/kg$ and $Re = 4 \times 10^4/m$

Next, the characteristics of separation and their variations with respect to $Ma$ are investigated, separation length, $L_{SC}/L_{RC}$, and separation angle. The results are shown in Fig. 37. With an increase in $Ma$, all models show a decrease in separation as well as $L_{SC}/L_{RC}$, while the latter of the perfect gas model remains larger than 1 and those of the real gas models are smaller than 1. The angles in Fig. 37(c) exceed 12.5° and show less variation overall, with equilibrium gas model demonstrating a slight increase with an increase in $Ma$.

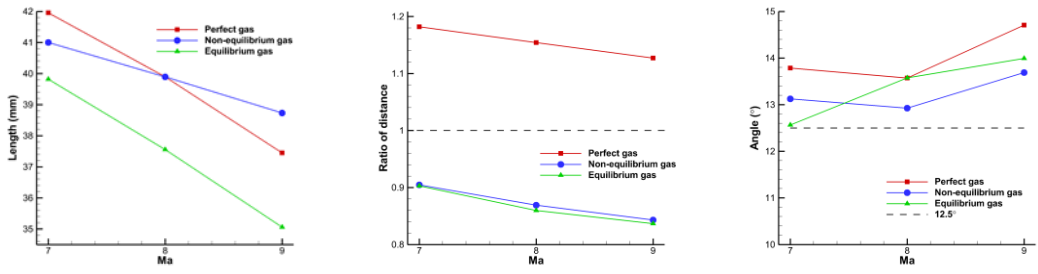

(a) Length of separation zone  (b) $L_{SC}/L_{RC}$  (c) Angle of separation streamline with respect to fore wedge

Fig. 37 Variation in geometric characteristics of separation zone with enthalpy of three gas models at $Ma = 7$ and $Re = 4 \times 10^4/m$

To reveal the thermodynamic effects of real gas models, the corresponding $\gamma$ distributions at $Ma = 9$ are shown in Fig. 38(a) and (b) for representation. The qualitative characteristics resemble the discussions in Section 4; specifically, $\gamma$ of the equilibrium gas model appears to be generally larger than that of the non-equilibrium gas model. In Fig. 38(c) and (d), the percentage contours of $(1.4 - \gamma_{NEG})/1.4$ at two $Ma$ are displayed. It is worth noting that, as shown in Fig. 27(d), there are no significant changes observed with an increase in $Ma$, and the distribution characteristics can be compared to those discussed in Section 4.

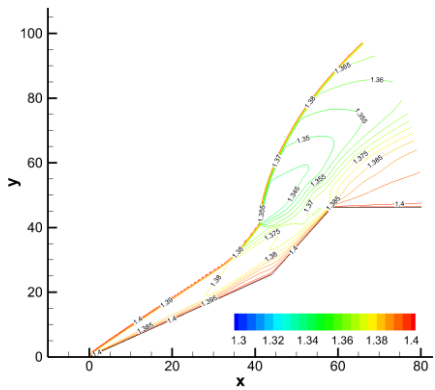
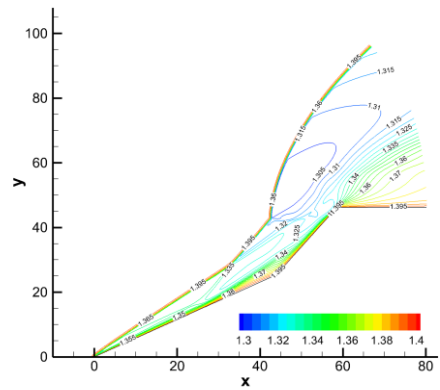

(a) $\gamma_{EG}$ at $Ma = 9$  (b) $\gamma_{NEG}$ at $Ma = 9$

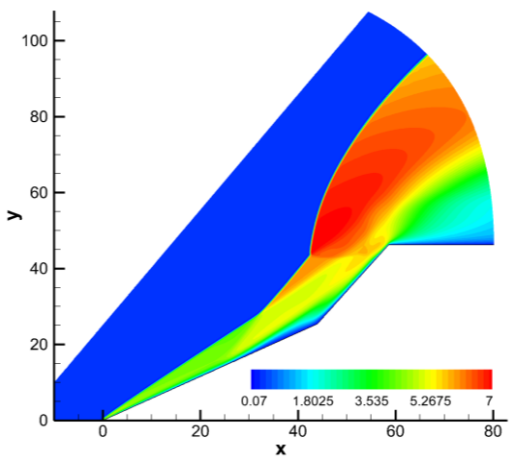
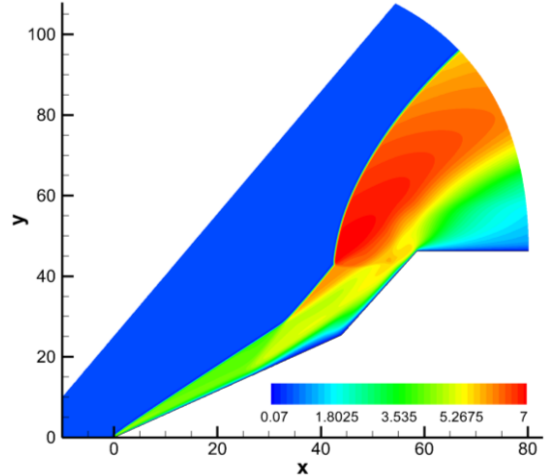

(c) $\frac{1.4-\gamma_{NEG}}{1.4} \times 100\%$ at $Ma = 8$      (d) $\frac{1.4-\gamma_{NEG}}{1.4} \times 100\%$ at $Ma = 9$

Fig. 38 Contours of specific heat ratio of two gas models at $Ma = 9$ as well as relative differences of the non-equilibrium gas model at $Ma = 8$ and 9 under $Re = 4 \times 10^4/m$ and $h_0 = 2.1 MJ/kg$

In terms of engineering considerations, Fig. 39 displays the variations of heat transfer and pressure coefficient with $Ma$. The figures reveal consistent patterns that have been previously discussed in Section 4. Specifically, the figures demonstrate that the heat flux decreases as $Ma$ increases on the fore wedge before the separation point. These decreases can be attributed to the reduction in inflow temperature, with similar trends appearing around 52–56 mm. The pressure distributions of the three models show nearly identical performance on the fore wedge before SS, aligning with their inviscid predictions. Similar distributions are observed after the expansion corner of the aft wedge, suggesting that the effect of $Ma$ is negligible in that region. However, a larger $Ma$ results in a higher pressure during separation and subsequent interaction with different degrees. Additionally, the distributions of heat transfer and pressure indicate that overall separation scale decreases with increasing $Ma$, as shown in Fig. 37(a).

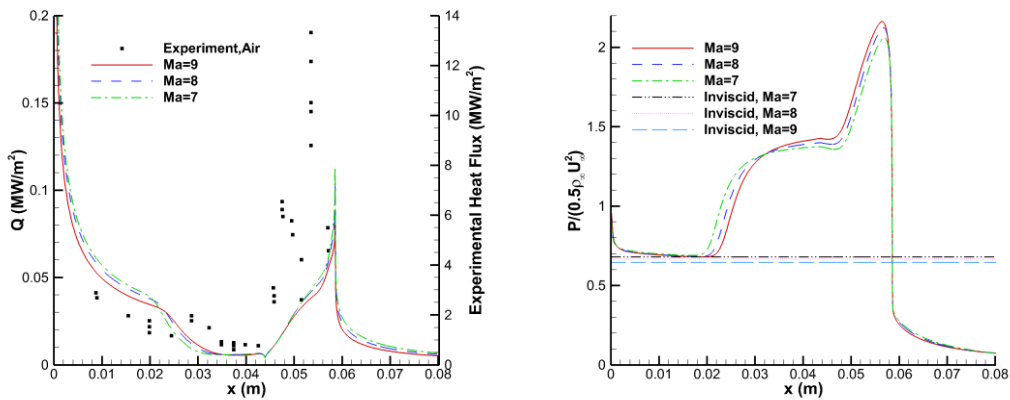

(a) Perfect gas

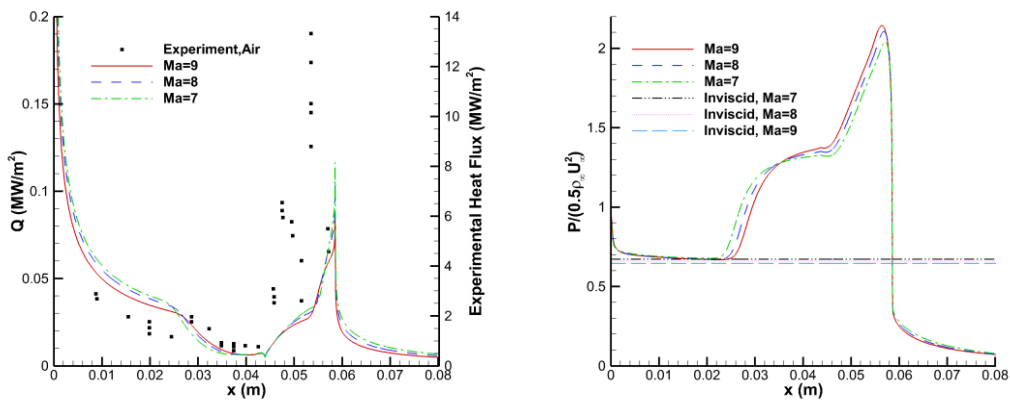

(b) Equilibrium gas

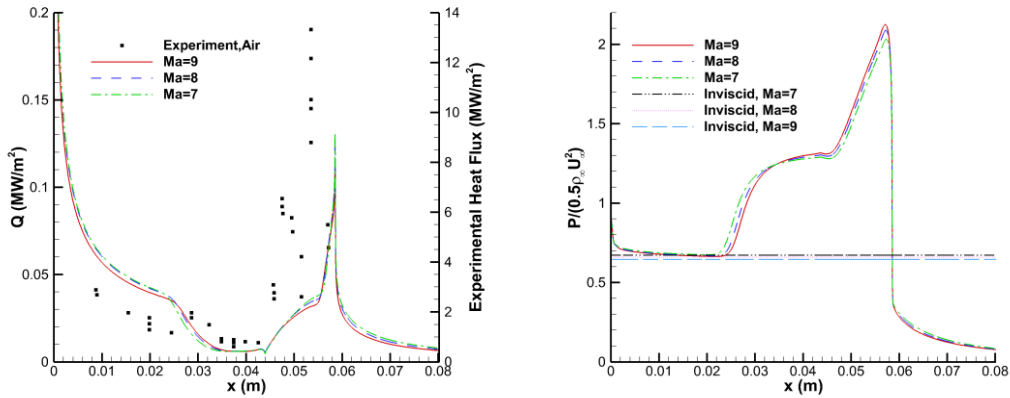

(c) Non-equilibrium gas

Fig. 39 Variations in aerodynamic properties with $Ma$ of three gas models at $Re = 4 \times 10^4/m$ and $h_0 = 2.1 MJ/kg$ with the reference of M7_2 [2, 3] for heat flux and inviscid prediction for pressure coefficient where the left column refers to heat transfer and the right refers to pressure coefficient

### 6.2  Effect of $h_0$ variation at $Ma = 7$ and $Re = 4 \times 10^4/m$

Given that $h_0 = 2.1MJ/kg$ in Section 4, the two additional $h_0$ considered here are chosen as $1.5MJ/kg$ and $2.5MJ/kg$. The other conditions are as follows: $\{T_\infty = 138.2K, p_\infty = 9.166pa\}$ and $\{T_\infty = 230.333K, p_\infty = 18.576pa\}$.

With the use of the three gas models, the computations are found to achieve steady solutions at the selected total enthalpies. However, the results obtained from the non-equilibrium gas model are used for discussion as a representative example. It is important to note that the analysis of the results at $h_0 = 2.1MJ/kg$ can be found in Section 4. Similar to the previous subsection, Fig. 40 shows the pressure, vortex structures, and temperature contours. Although there were no fundamental changes observed with an increase in $h_0$, there was a noticeable enhancement in the strength of SS and TS. Additionally, minor variations in separation occurred, such as downstream movement and reattachment of separation.

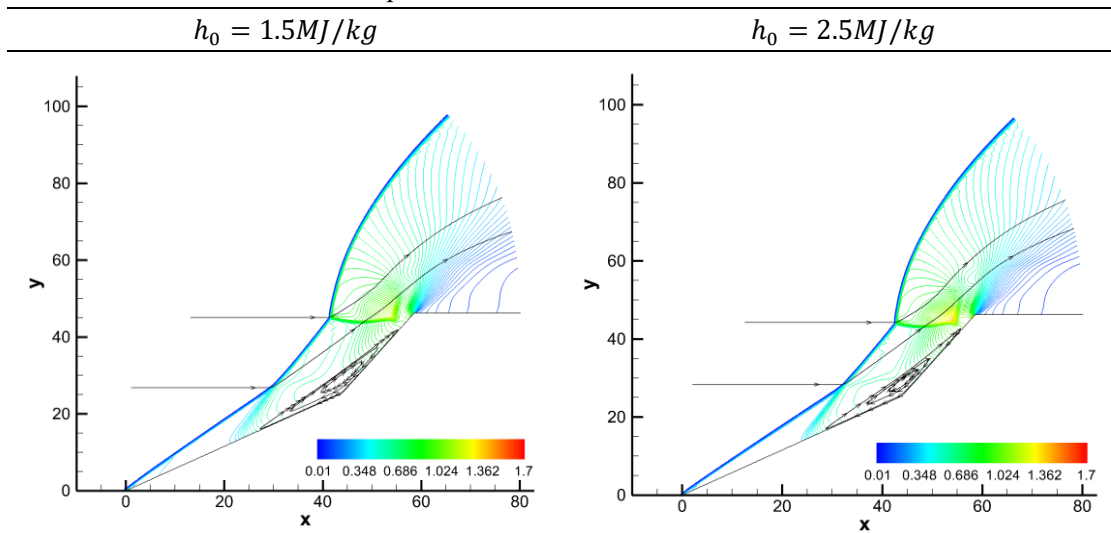

(a)  Pressure contours with two slip lines and vortex shown by streamlines

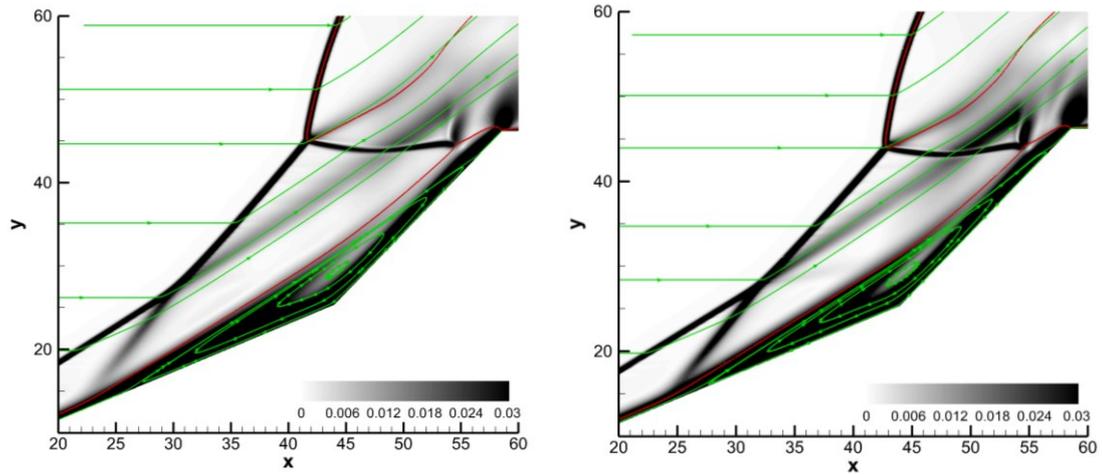

(b) Vortex structure with background of density-gradient and sonic lines

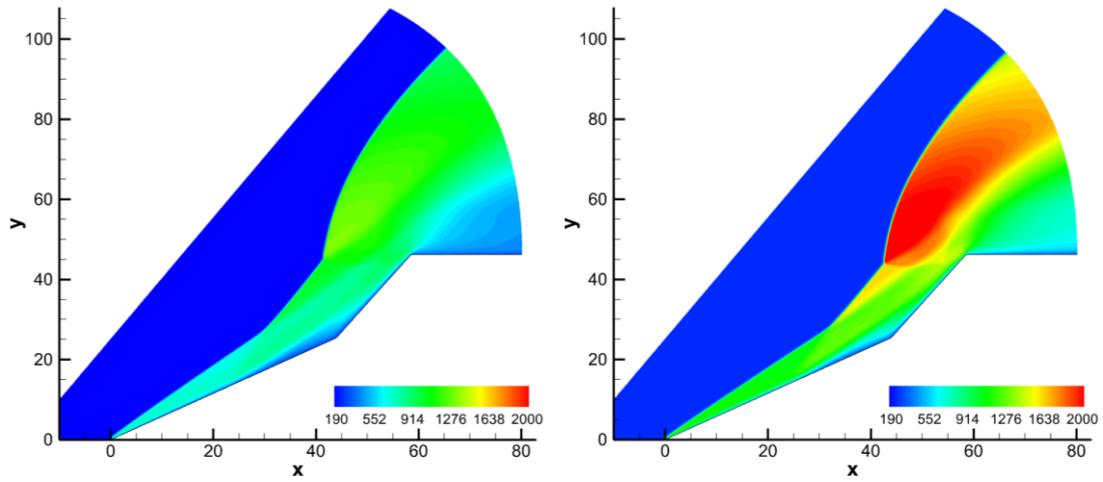

(c) Temperature contours

Fig. 40 Variations in variable contour and vortex structure with enthalpy at $Ma = 7$ and $Re = 4 \times 10^4 /m$

Similarly, the trajectories of impingement and triple points of the non-equilibrium gas model are examined and shown in Fig. 41. Similar coordinate transformations have been applied, along with arrows indicating the direction of $h_0$ increase. The figure illustrates that as $h_0$ increases, the impingement points of the real gas models exhibit a slight movement towards the expansion corner, while those of the perfect gas model show a subdued departure. Meanwhile, the triple point of the perfect gas model demonstrates a diminishing upstream movement, whereas those of the real gas models tend to converge in approximately opposite directions.

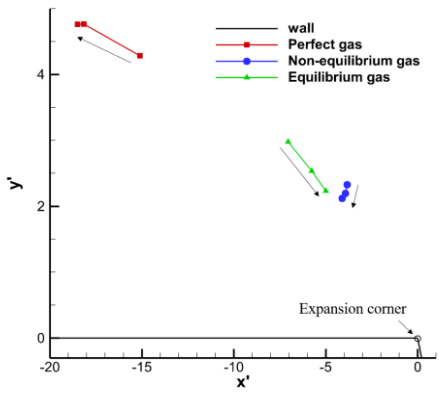
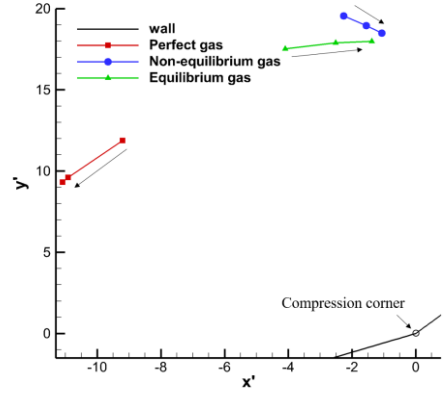

(a) Impingement point  (b) Triple point

Fig. 41 Trajectories of characteristic points of three gas models with $h_0$ at $Ma = 7$ and $Re = 4 \times 10^4/m$

Next, the separation length, $L_{SC}/L_{RC}$, separation angle, and their variations with respect to $h_0$ are measured and shown in Fig. 42. It can be observed that as $h_0$ increases, the length of the separation zone in the perfect gas model increases, while those in the real gas models remain relatively stable. Additionally, $L_{SC}/L_{RC}$ of the latter decreases relatively compared to the former, while their values relative to one remain unchanged. The angles in Fig. 42(c) are consistently above 12.5°, showing no significant change.

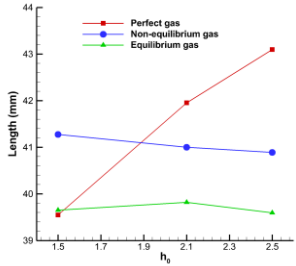 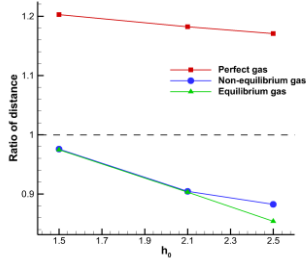 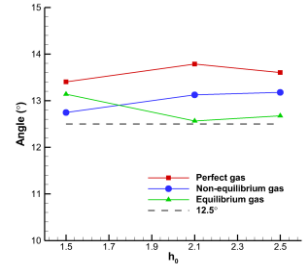

(a) Length of separation zone    (b) $L_{SC}/L_{RC}$    (c) Angle of separation streamline with respect to fore wedge

Fig. 42 Variation in geometric characteristics of separation zone with $h_0$ of three gas models at $Ma = 7$ and $Re = 4 \times 10^4/m$

To illustrate the effect of gas models on thermodynamics, contours of the percentage of $(1.4 - \gamma)/1.4$ of real gas models at $h_0 = 1.5MJ/kg$ and $2.5MJ/kg$ are shown in Fig. 43, while those of the non-equilibrium gas model have been shown in Fig. 27(d). The figures reveal that, for the equilibrium gas model, $\gamma$ is predominantly less than 1.4 in most areas, as discussed in Section 4. However, a region with $\gamma > 1.4$ emerges near the compression corner, close to the wall, and after the expansion corner, as depicted by the white line $\gamma = 1.4$ in Fig. 43(a). In contrast, $\gamma$ remains below 1.4 throughout for the non-equilibrium gas model, as shown in Fig. 43(b). The distributions and corresponding variations within these models differ noticeably, e.g. there is a notable deviation from 1.4 after TS near the impingement with further deviation as $h_0$ increases. Such deviations suggest different performances among real gas models.

| $h_0 = 1.5MJ/kg$ | $h_0 = 2.5MJ/kg$ |
| --- | --- |

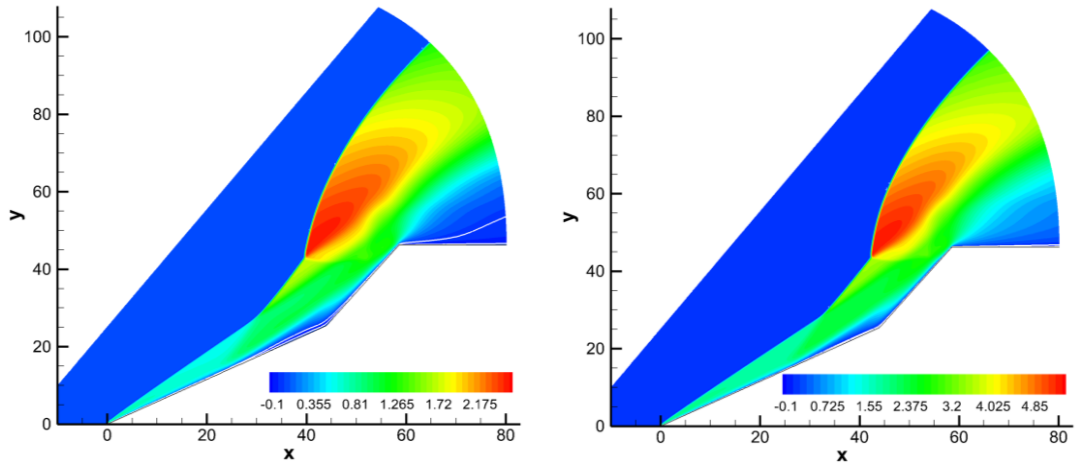

(a) Equilibrium gas model with the white line denoting $\gamma = 1.4$

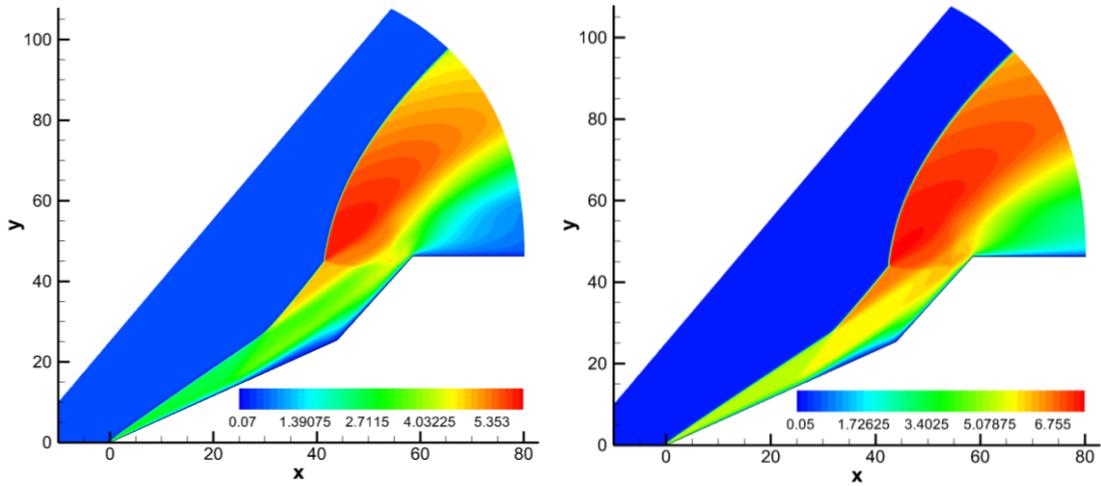

(b) Non-equilibrium gas model

Fig. 43 Contours of $(1.4 - \gamma)/1.4 \times 100\%$ of two gas models at $h_0 = 1.5 \ and \ 2.5 \ MJ/kg$ under $Ma = 7$ and $Re = 4 \times 10^4/m$

Finally, the heat transfer and pressure coefficient distributions at three $h_0$ are shown in Fig. 44. The heat transfer results suggest that, although the distribution trends are similar for all three models, a higher heat flux is observed with a higher $h_0$. The differences are less apparent in the separation region ahead of the wedge deflection. Additionally, other characteristics are similar to those discussed in Section 4. In contrast, the pressure coefficients show close distributions, with the exception of relatively higher dissipation observed in the case of $h_0 = 1.5 MJ/kg$ around SS. Therefore, the current variations in $h_0$ have a lesser effect on the pressure coefficient compared to the effect on heat transfer.

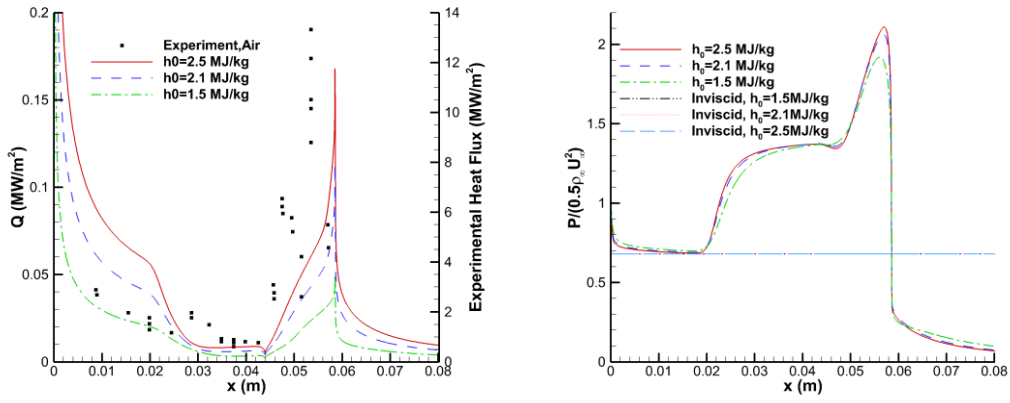

(a) Perfect gas

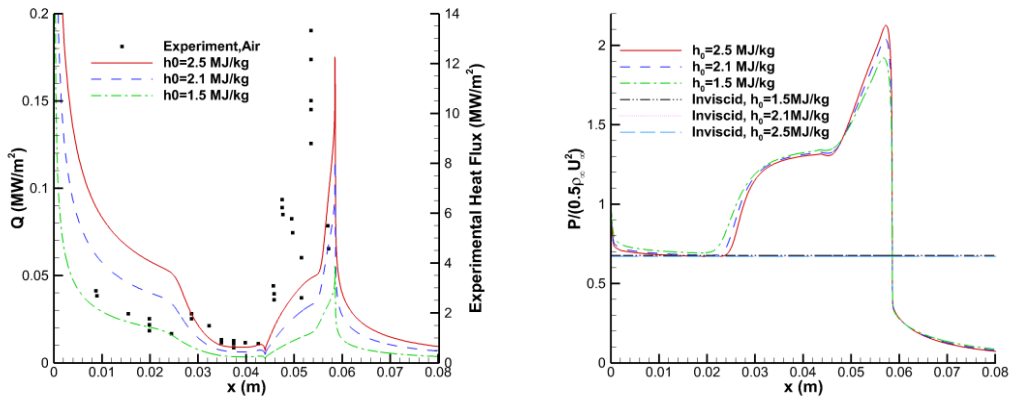

(b) Equilibrium gas

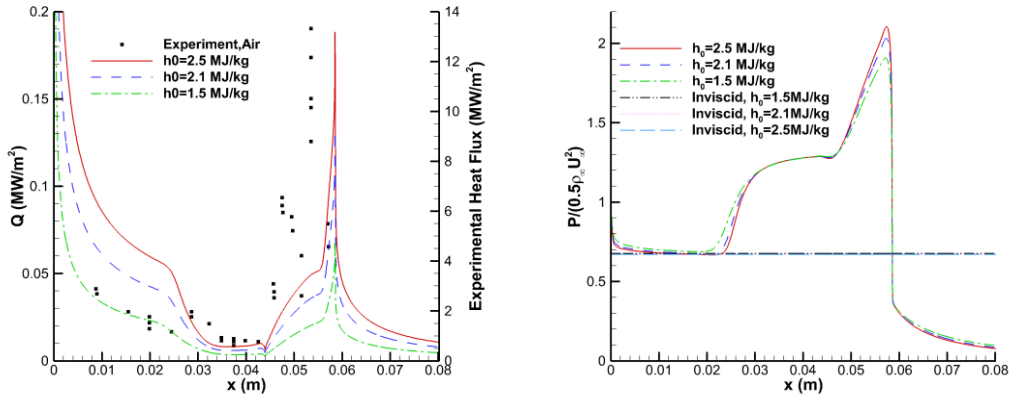

(c) Non-equilibrium gas

Fig. 44 Variations in aerodynamic properties with $h_0$ of three gas models at $Ma = 7$ and $Re = 4 \times 10^4/m$ with the reference of M7_2 [2, 3] for heat flux and inviscid prediction for pressure coefficient where the left column refers to heat transfer and the right refers to pressure coefficient

## 7  Conclusions

Regarding the steady interaction of the hypersonic 30–55° double wedge, comprehensive investigations were carried out around the condition of M7_2 as proposed by Swantek & Austin [2, 3], but at lower $Re$ using three gas models. Subsequently, changes in the interactions were studied by alternatively changing $Ma$ and $h_0$. In addition to simulations, analytic studies were conducted using the shock polar method and the 1D model of isentropic flows and shock relations. The following conclusions were drawn:

(1) In the referenced case M7_2 ($Ma = 7$ and $h_0 = 2.1 MJ/kg$), the upper limits of $Re$ were determined for computations to achieve steady interactions, leading to the development of a diagram illustrating flow steadiness versus unsteadiness. As $Re$ approaches the limits, the impingement of the TS will approach the expansion corner and converge at certain locations.

(a) At the upper limits of $Re$, a comprehensive analysis of steady interactions involving LS, SS, CS, BS, TS, etc. was conducted, and detailed characteristics were identified. General features, such as the larger interaction size of the perfect gas model compared to real gas models, were observed. The interaction of the former exhibited additional complexities, such as QNS and a corresponding subsonic zone within the slip line passage. Meanwhile, $\gamma$ distributions of the latter showed deviations from 1.4, indicating lower temperature and implying the potential influence of the real gas model.

(b) As $Re$ decreased from $4 \times 10^4/m$, variations in steady interactions were observed, along with detailed characteristics. Despite a general reduction in interaction size, a specific pattern of the perfect gas model was noted in which the TS impinged and reflected over the separation bubble, while those of the real gas models were still preserved. The geometric characteristics of triple points indicated a converging trend with decreasing $Re$, whereas those of separation continued to evolve, including changes in size and angle of separation, $L_{SC}/L_{RC}$ etc. In terms of gas models, a unique feature emerged, i.e., $L_{SC}/L_{RC} > 1$ for the perfect gas case while $L_{SC}/L_{RC} < 1$ for the real gas cases. Heat transfers of the three models indicated the same changes with decreasing $Re$, while the pressure coefficients showed distributions consistent with inviscid predictions before SS and with mutual differences after separation.

(c) Flow models were proposed by simplifying the abovementioned steady interactions. To understand the mechanism, the inviscid polar method, either for the perfect gas or equilibrium gas model, was applied to analyze the interactions at $Ma = 7$ and $h_0 = 2.1 MJ/kg$ with varying $Re$. Comparisons with the computations generally showed reasonable agreement, except for one case with the perfect gas model in the region after the TS. Meanwhile, a 1D flow model was used to elucidate the mechanism of QNS in the perfect gas model at $Re = 9.5 \times 10^4/m$, $Ma = 7$, and $h_0 = 2.1 MJ/kg$, providing a reasonable prediction of the occurrence of QNS.

(2) By changing $Ma$ and $h_0$ alternatively around the condition $Ma = 7$, $h_0 = 2.1 MJ/kg$ at $Re = 4 \times 10^4/m$, the effects of their variations were understood as follows: (a) An increase in $Ma$ led to a decrease in interaction size, particularly in heat transfer. The pressure coefficients of different models showed generally similar distributions with slight differences. (b) An increase in $h_0$ resulted in similar changes in the separation zone of the perfect gas model, while fewer variations were observed in the real gas cases. In flows of the equilibrium gas model, a region with $\gamma > 1$ existed, whereas in the non-equilibrium case, $\gamma$ was always less than 1.

Despite the aforementioned efforts and insights, differences in the interactions of the three models are evident at low enthalpies. Currently, there is not enough confidence to determine which gas model would yield more accurate results. Further investigation, particularly through experimental means, is necessary for clarification.